\documentclass[aps, nofootinbib, prd, superscriptaddress, tightenlines,
notitlepage, twocolumn, floatfix]{revtex4-1} %linenumbers
\usepackage{tabularx}
\usepackage{graphicx}
\usepackage{amssymb,bm,tensor}
\usepackage{textcomp}
\usepackage{xcolor}
\usepackage{amsmath}
\usepackage[varg]{txfonts}
\usepackage{enumerate}
\usepackage{mathtools}
\usepackage[normalem]{ulem}
\usepackage{bm} % bold math
\usepackage[OT1]{fontenc}
\usepackage[colorlinks]{hyperref}

\newcommand{\LS}[1]{{ #1}} % Lijing Shao
\newcommand{\KIAA}{\affiliation{Kavli Institute for Astronomy and
Astrophysics, Peking University, Beijing 100871, China}}
\newcommand{\DOA}{\affiliation{Department of Astronomy, School of Physics,
Peking University, Beijing 100871, China}}

\newcommand{\SOP}{\affiliation{School of Physics and State Key Laboratory
of Nuclear Physics and Technology, Peking University, Beijing 100871,
China}}
\newcommand{\NAOC}{\affiliation{National Astronomical Observatories,
Chinese 
Academy of Sciences, Beijing 100012, China}}
\newcommand{\BNU}{\affiliation{Department of Astronomy, Beijing Normal
University, Beijing 100875, China}}

\begin{document}

\title{Improved deep learning techniques in gravitational-wave data
analysis}
\date{\today}
\author{Heming Xia}\DOA%\CS
\author{Lijing Shao}\email{lshao@pku.edu.cn}\KIAA\NAOC
\author{Junjie Zhao}\SOP
\author{Zhoujian Cao}\BNU

\begin{abstract} 
In recent years, convolutional neural network (CNN) and other deep learning
models have been gradually introduced into the area of gravitational-wave
(GW) data processing. Compared with the traditional matched-filtering
techniques, CNN has significant advantages in efficiency in GW
signal detection tasks. In addition, matched-filtering techniques are based
on the template bank of the existing theoretical waveform, which makes it
difficult to find GW signals beyond theoretical expectation. In this paper,
based on the task of GW detection of binary black holes, we introduce the
optimization techniques of deep learning, such as batch normalization and
dropout, to CNN models. Detailed studies of model performance are carried
out. Through this study, we recommend to use batch normalization and
dropout techniques in CNN models in GW signal detection tasks. Furthermore,
we investigate the generalization ability of CNN models on different
parameter ranges of GW signals. We point out that CNN models are robust to
the variation of the parameter range of the GW waveform. This is a major
advantage of deep learning models over matched-filtering techniques.
\end{abstract}

\maketitle

%% main text
%---------------------------------------------------------------------
\section{Introduction}
\label{sec:intro}
%---------------------------------------------------------------------

On 11 February 2016, LIGO and Virgo Collaboration announced the first
detection of gravitational-wave (GW) signals from 14 September 2015, the
so-called GW150914~\cite{Abbott:2016blz, Abbott:2017xlt, Liu:2016hxz}. The
detected GW signal comes from a binary black hole (BBH) merger. The masses
of the BBH are estimated to be 29\,$M_{\odot}$ and 36\,$M_{\odot}$. The
successful observation of GW signals has provided valuable experimental
data for GW astronomy, setting off a wave of GW researches. So far, 50 GW
signals from compact binary coalescences have been successfully
detected~\cite{LIGOScientific:2018mvr, Nitz:2018imz,
LIGOScientific:2020stg, Abbott:2020khf, Abbott:2020uma, Abbott:2020tfl,
Abbott:2020niy}. Except for the binary neutron star (BNS) merger event,
GW170817~\cite{TheLIGOScientific:2017qsa} and three other {\it
strictly-speaking} unclear merger events---GW190814~\cite{Abbott:2020khf},
GW190425~\cite{Abbott:2020uma}, and
GW190426\_152155~\cite{Abbott:2020niy}---the remaining 46 signals all come
from BBH mergers.

Currently, both LIGO and Virgo mainly use the matched-filtering
method~\cite{Finn:1992wt, Cannon:2011vi, Usman:2015kfa,
TheLIGOScientific:2016uux} to detect GW signals. This method builds a
theoretical waveform template bank to match the monitored data and captures
trigger signals as candidates for further verification~\cite{Finn:1992wt}.
Matched filtering plays a vital role in the processing of GW signal
detection. However, it has shortcomings that should not be
overlooked~\cite{Smith:2016qas, Harry:2016ijz}. Matched-filtering method
requires a full search in the template bank to match the signal, which
limits the data processing speed and, if a big template bank is used, has
difficulty to meet the needs of real-time observation. In addition, with
continuous expansion of theoretical waveforms in an enlarging parameter
space, the search space of matched filtering increases, which leads to an
increase of data processing time and a reduction in the processing
speed~\cite{Harry:2016ijz}.

In recent years, many machine learning methods have been developed in GW
signal detection tasks~\cite{Cornish:2014kda, Chua:2018woh, Chua:2019wwt,
Mukund:2016thr, Colgan:2019lyo, Cuoco:2020ogp}. In the machine learning
field, as AlexNet won the championship in the ImageNet competition in
2012~\cite{Krizhevsky:2017misc}, deep learning algorithms stood out and
achieved great success in many fields such as image classification, natural
language processing, and speech recognition~\cite{schmidhuber:2015,
Goodfellow:2016}. In terms of classification tasks, compared with
traditional machine learning algorithms, many deep learning algorithms,
including convolutional neural networks (CNNs), have made significant
progress in model accuracy and model complexity. Besides, the
characteristics of the deep learning algorithm make the time-consuming training
process be completed offline before the actual data analysis. It greatly
reduces the amount of calculation in the online process and meets the need
of real-time detection~\cite{Goodfellow:2016}.

At present, deep learning methods, especially CNNs, have been widely
explored in GW data processing~\cite{George:2016hay, Gabbard:2017lja,
Krastev:2019koe, Gabbard:2019rde,Shen:2019vep, Shen:2017jkj,
Chatterjee:2019gqr,Dreissigacker:2019edy, Gebhard:2019ldz}. In 2017, George
and Huerta~\cite{George:2016hay} firstly applied CNN to GW signal detection
tasks. They generated mock GW signals from BBH mergers and added them into
white Gaussian noise to generate simulation datasets. They pointed out that
the sensitivity, namely the fraction of signals which are correctly
identified, of CNNs is similar to the matched-filtering method, while its
speed has been greatly improved. The work in the same period by
\citet{Gabbard:2017lja} compared the false alarm rate and the receiver
operating characteristic (ROC) curve between CNN models and the
matched-filtering method, leading to a similar conclusion. Subsequently,
the application of deep learning in the field of GW signal detection has
been expanded greatly. \citet{Krastev:2019koe} indicates that deep learning
models work better on BNS mergers than BBH mergers. More deep learning
models such as residual network, fully CNN, and other structures have been
introduced~\cite{Dreissigacker:2019edy, Gebhard:2019ldz, George:2018awu,
Li:2020kgf, Marulanda:2020nww, Wang:2019zaj}, and the research field has
been continuously expanding. Variational autoencoders and Bayesian neural
networks are used for parameter estimation of GW
signals~\cite{Gabbard:2019rde, Shen:2019vep}. Long short-term memory
network has made progress in the field of GW signal noise reduction, which
proves that it can effectively remove environmental noise and restore the
GW signal under noise~\cite{Shen:2017jkj}. In the sky localization
searching task of GW signals, deep learning methods such as CNNs have also
achieved good results~\cite{Chatterjee:2019gqr}.

However, almost all deep learning algorithms such as CNNs used in the
current researches are basic models. It means that they can be further
optimized. In addition, many studies have pointed out that deep learning
models can maintain a certain degree of robustness to GW signals beyond
the range of the training set parameters~\cite{George:2016hay, Gabbard:2017lja,
Gebhard:2019ldz}. However, there is no specific research on this aspect.
Grounded on the above two points, we conduct experiments on the
optimization effects of several deep learning techniques in the field of GW
signal detection. The result shows that, compared with the basic model, the
model with improved techniques achieves better performance. On the low
signal-to-noise ratio (SNR) dataset, the model with multiple improved
techniques has an accuracy rate of $84\%$ on the testing set, $12\%$ higher
than that of the basic model, and an area under curve (AUC) score of 0.91,
$6\%$ higher than that of the basic model. On the overall dataset, our
model with multiple improved techniques has an accuracy rate of $94\%$ on
the testing set, $4\%$ higher than that of the basic model, and an AUC
score of 0.98, $2\%$ higher than that of the basic model. Moreover, we make
a detailed research on the robustness of CNN models on GW signal detection
tasks. Our experiments show that the CNN model has good robustness for data
of different parameter ranges for masses and spins.

This paper is organized as follows. In Sec.~\ref{sec:deep learning}, we
give a brief overview on deep learning. In Sec.~\ref{sec:set}, we introduce
our simulated dataset to be used in our experiments. Then the improved
techniques for CNN and the corresponding experimental results are shown in
Secs.~\ref{sec:techniques} and~\ref{sec:res}, respectively. In
Sec.~\ref{sec:robust}, we investigate the generalization ability of the CNN
model in different parameter ranges.

%---------------------------------------------------------------------
\section{Deep Learning}
\label{sec:deep learning}

Traditional machine learning methods include $k$-nearest neighbor, decision
tree, support vector machine, and so on~\cite{Mitchel:1997}. The advantage
of machine learning methods is that to some extent, they can replace the
process of human learning. Through training on a large dataset, theses
models can learn the {\it relationship} between data so that to classify,
predict, and help human to make decisions~\cite{bishop:2006}. However, when
traditional machine learning methods are applied to specific tasks, they
usually have difficulty processing the original data. In some cases,
researchers have to manually extract data features and put them into
algorithms. Besides, traditional machine learning methods are usually
limited by their fixed model structure. It is difficult for these
algorithms to achieve rapid improvement in computing power and
accuracy~\cite{Goodfellow:2016}.

Deep learning overcomes some limitations of traditional machine learning
methods. Its algorithm is derived from neural network which is a sub-field
of traditional machine learning. This algorithm solves the limitation of
the depth of neural network and increases the computational power of the
model. So far, many deep learning models such as CNN, residual network, and
long and short time memory have been proposed. At present, deep learning
has achieved substantial success in face recognition, automatic driving,
speech processing, and many other fields~\cite{Goodfellow:2016}. For the
sake of a self-contained work, below we briefly review the principle of
deep learning, including the structure of neurons, the basic principle of
neural network, and the model of CNN.

%---------------------------------------------------------------------
\subsection{Neuron}
\label{subsec:neuron}
%---------------------------------------------------------------------

\begin{figure}%[htbp!]
		\centering
		\includegraphics[width = 8cm]{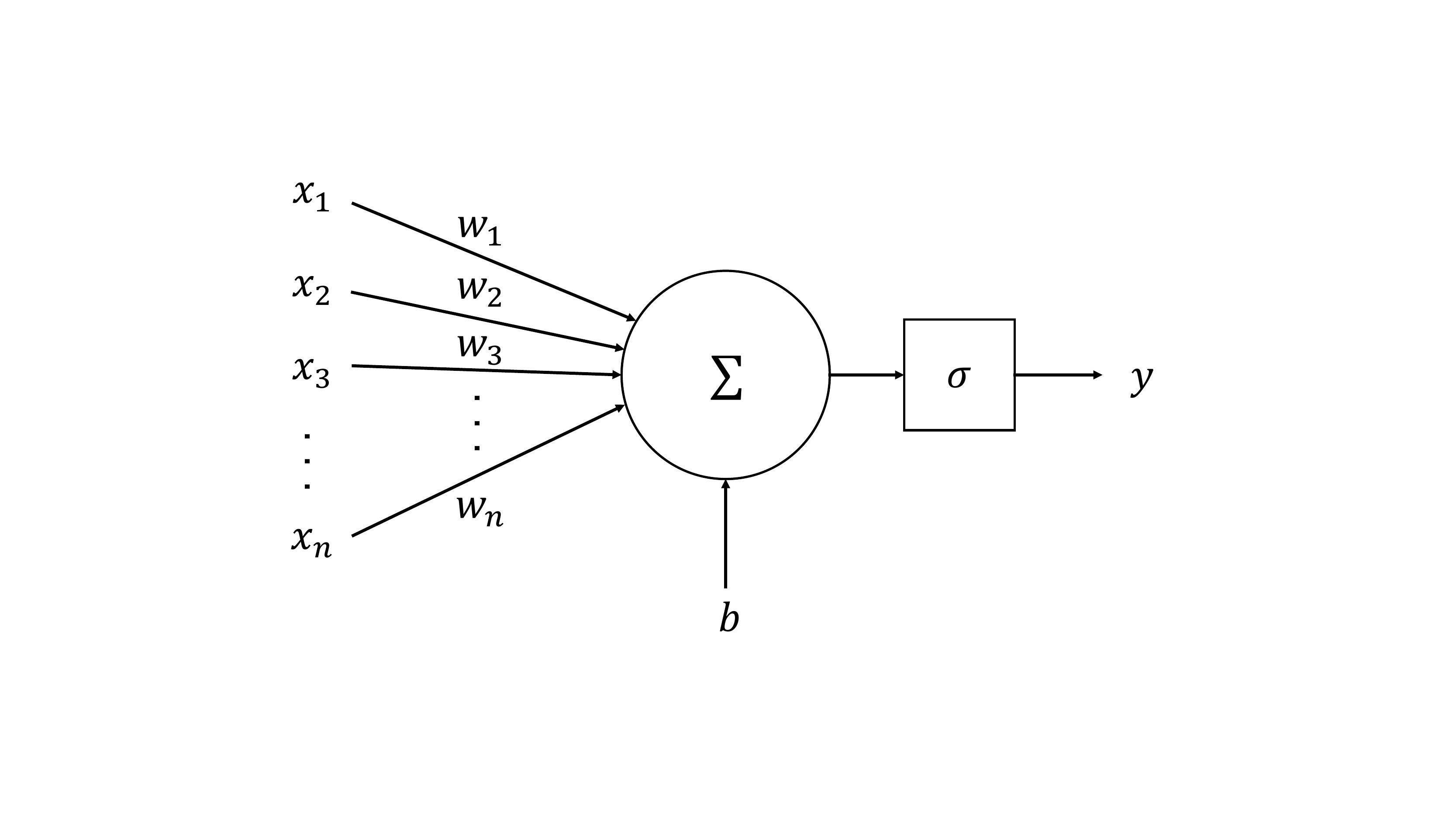}
		\caption{Architecture of a neuron.}
		\label{fig:neuron:architecture}
\end{figure}

The idea of neural networks in machine learning evolved from biological
models. Generally speaking, neural network is a network of parallel
interconnections composed of simple adaptive units~\cite{Kohonen:1988}. Its
organization can simulate the interaction of the biological nervous system
to real-world objects. The neuron is the ``simple unit'' in the above
definition. In 1943, \citet{McCulloch:1943} abstracted it into the simple
model shown in Fig.~\ref{fig:neuron:architecture}, namely the ``M-P neuron
model''. In this model, each neuron receives input data $\boldsymbol
x=(x_1,x_2,...,x_n)$ from previous neurons. The input is multiplied by the
weight $\boldsymbol w=(w_1,w_2,...,w_n)$, plus the bias $b$, and then is
injected into the activation function $\sigma$ to obtain the output $y$. A
single neuron can be represented by the following formula using vectors,
%--
\begin{equation}
	y=\sigma(\boldsymbol w\boldsymbol   x^\intercal + b) \,.
\end{equation}
%--

The activation function $\sigma$ introduces nonlinear operations into neurons.
Otherwise, the structure of neurons will simply be a superposition of
linear operations, and the power of the network will be greatly reduced.
Activation function $\sigma$ comes in many forms~\cite{Nwankpa:2018}. In
this paper, we use the most commonly used activation function
{\tt ReLU}~\cite{Nair:2010} in our study,
%--
\begin{equation}
	\label{eq:ReLU}
	\sigma(x)=
	\begin{cases}
		x \,, & x>0 \,,\\ 
		0 \,, & x\leq0 \,.
	\end{cases}
\end{equation}
%--
In general, the current neuron will only output positive values after
calculating the weighted sum of data from the first $N$ neurons. From the
feature level, it can be understood as the following. After a linear
combination of $N$ features, neurons input the combined features into the
activation function to obtain the output features.

%---------------------------------------------------------------------
\subsection{Neural Network}
\label{subsec:nn}
%---------------------------------------------------------------------

% \subsubsection{Neural Network Architecture}

\begin{figure}%[htbp!]
		\centering
		\includegraphics[width = 8.5cm]{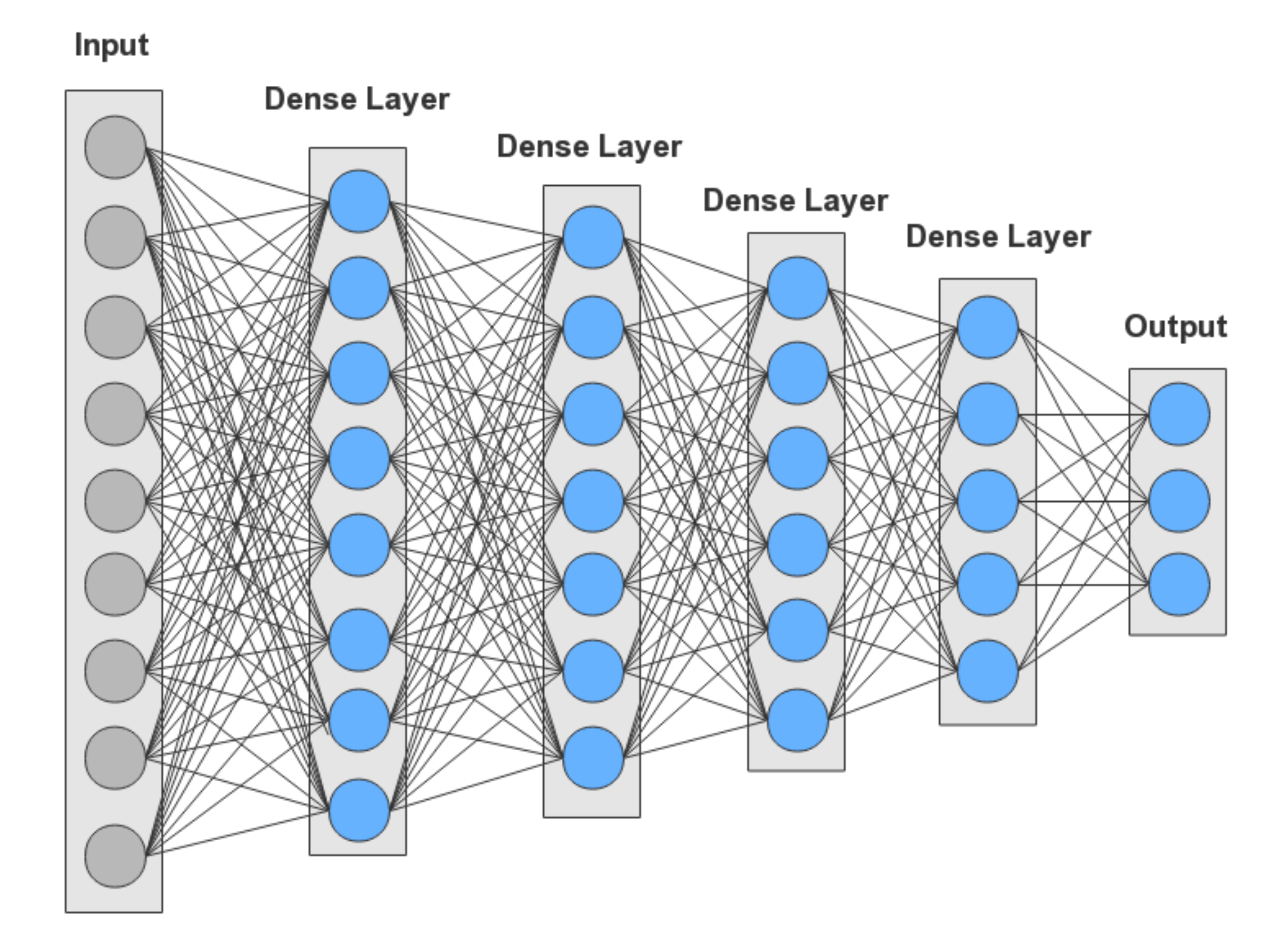}
		\caption{Structure of a FCNN.}\label{fig:neural:network}
\end{figure}

The simplest example of neural network is Fully Connected Neural Network
(FCNN). The structure of FCNN is shown in Fig.~\ref{fig:neural:network}. A
manually specified number of neurons constitute each layer, which is called
a ``Dense Layer'' of the network. Neurons between different layers are
independent. Each neuron receives data from the former layer, calculates
the result, and puts them into all neurons of the next
layer~\cite{Hopfield:1982pe}.
%%
% After the network receives the input data, the first layer directly
% analyzes the data, captures the features, and puts them into the second
% layer. The second layer performs further analysis and combines these
% features. This process is repeated until the final layer gets the output.
%%
Note that the input data size has a linear relationship with the parameters
of the first layer in FCNN~\cite{Goodfellow:2016}. With the input data size
increases, the number of parameters in the network will increase
correspondingly, slowing down the learning speed of the model and
increasing the requirement of data storage~\cite{schmidhuber:2015}.

% \subsubsection{Learning Strategy}

Neural network is a supervised learning algorithm, whose characteristic is
to use the labeled data---which is the correct output $y$---for training,
and test the model on unlabeled data. A loss function is defined to measure
the difference between the model output $\hat y$ and the correct $y$. The
expectation for training is to make the value of loss function as small as
possible~\cite{Goodfellow:2016}. The parameters to be informed in the
neural network are the weights $\boldsymbol w$'s and biases $b$'s in each
layer of neurons. According to the gradient descent strategy, parameters in
the neural network are updated to the direction where the value of loss
function decreases~\cite{sra:2012}. The gradient descent strategy reads,
%--
\begin{equation}
	\theta^{\rm new}= \theta^{\rm
	old}-\alpha\triangledown_{\theta}J(\theta) \,,
\end{equation}
%--
where $\theta$ is the parameter to be updated, $J$ is the loss function,
and $\alpha$ is the manually specified learning rate which controls the
speed of parameter updates.

% \subsubsection{The Overfitting Problem}

The learning of neural network is based on the training dataset collected
and annotated by human beings. However, when we finally apply the model to
real tasks, we hope the neural network to have good {\it generalization
ability}. {\it Generalization ability} with respect to the neural network
is defined as the ability of the network to handle unseen
patterns~\cite{uro:2011}. In other words, this concept measures how
accurately an algorithm is able to predict outcome values for previously
unseen data~\cite{mohri:2018}. The testing set is a good type of unseen
data. The data characteristics of the testing set are similar to the
training set. In the meantime, its distribution is independent of the
training set, and it does not appear in the training
process~\cite{Goodfellow:2016}. Therefore, the final result of the testing
set is a reliable index to evaluate the performance of the neural network.

However, if the number of parameters in the network is more compared to the
samples in the training set, {\it overfitting} occurs~\cite{lawrence:2000}.
If there are too many parameters in the neural network, the predicting
power of the model will get too strong, which makes the model fit the
(noisy) characteristics of the training dataset too much in the learning
process. As a result, the model is only effective for the data samples that
appear in the training set, resulting in the so-called {\it the overfitting
problem}~\cite{hawkins:2004}. Therefore, how to design the model structure
of neural network with appropriate layers and neuron numbers with strong
data generalization ability is an outstanding issue.

\subsection{Convolutional Neural Network}
\label{subsec:cnn}

\begin{figure}%[htbp!]
		\centering
		\includegraphics[width = 8.5cm]{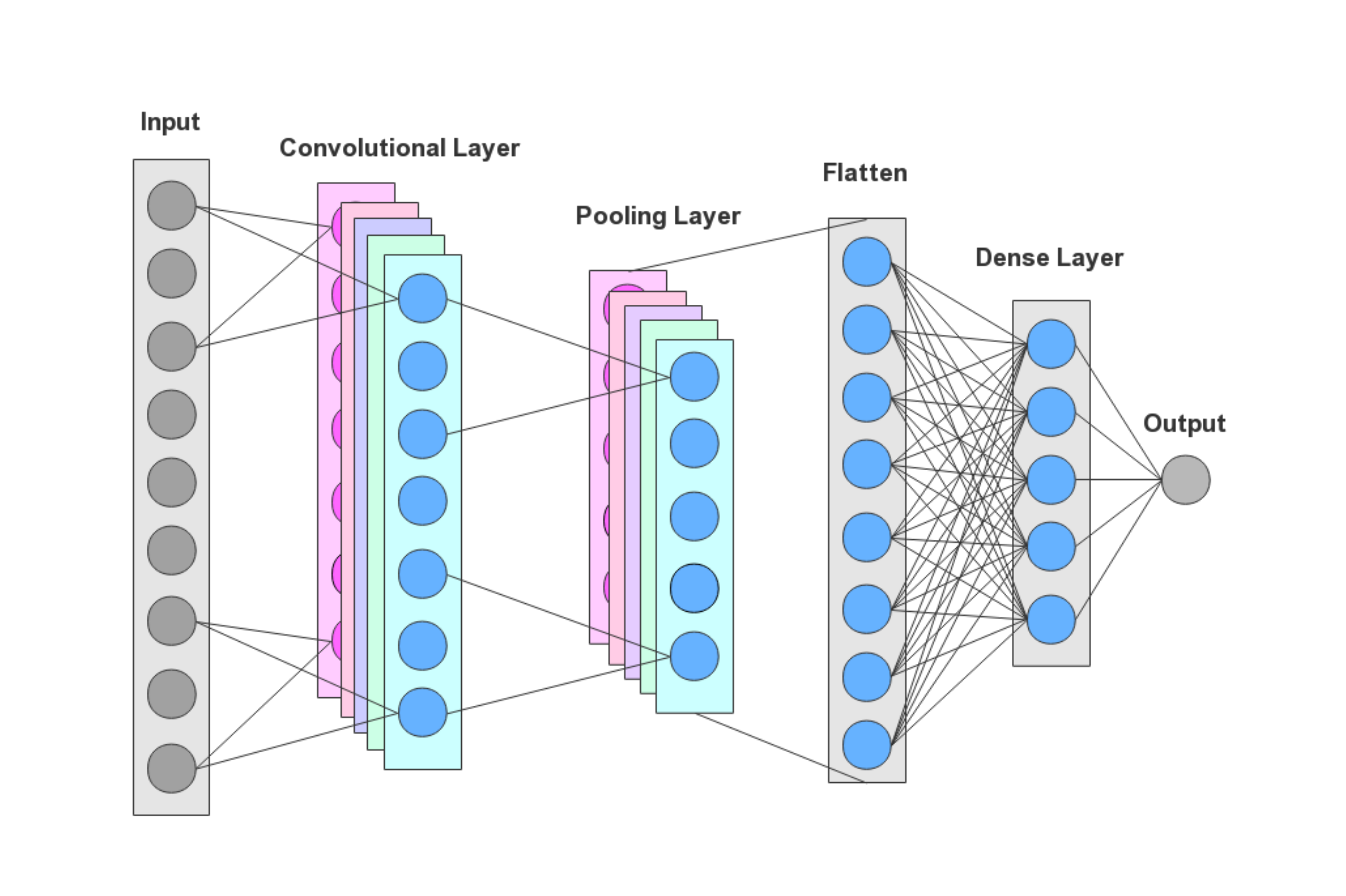}
		\caption{Structure of a CNN.}\label{fig:convolutional:neural:network}
\end{figure}

As shown in Fig.~\ref{fig:convolutional:neural:network}, the structure of
CNN is divided into the convolutional layer, the pooling layer, and the
dense layer~\cite{Goodfellow:2016}. Each convolutional layer is composed of
a specified number of kernels. Each kernel multiplies the input feature
values with weights, and adds the biases to obtain outputs. Different
kernels get different parameter values after training. The pooling layer
itself does not contain any parameters. Take the max-pooling layer as an
example. After receiving the input data, this layer scans the data
according to a specified stride within a window of a certain length. Then,
it outputs the maximum value of the data in each scanning
window~\cite{Yamaguchi:1990}. Therefore, the pooling layer compresses the
data. It checks all the features in the scanning window and chooses the
most important one~\cite{ciresan:2012}. There are other pooling methods,
e.g., the average pooling, which outputs the average value of the data in
the window~\cite{Aghdam:2017}.

The pooling layer plays an important role in improving the receptive field
of CNN. After the data passes through the pooling layer, the original data
length gets shortened, and the kernel in the next convolutional layer can
handle a larger range of data than the convolution window in the previous
layer, thereby it expands the convolutional layer's overall operation range
of the data~\cite{Krizhevsky:2017misc}. After extracting the features of
the input data through convolutional layers and pooling layers, the
flattened feature map will be put into the fully connected layer to obtain
the final output~\cite{Goodfellow:2016}.

Since the parameters to be learned by the convolutional layer are only the
parameter values in the convolution kernel, they are independent of the
input data size. Therefore, CNN reduces the number of free parameters,
allowing the network to be deeper with fewer parameters~\cite{Aghdam:2017}.
Due to its unique model structure and powerful data processing capability,
CNN is being widely used~\cite{Goodfellow:2016}.

%---------------------------------------------------------------------
\section{Simulated Dataset}
\label{sec:set}
%---------------------------------------------------------------------

In this section we discuss our strategy to simulate GW data and build
different datasets for machine learning studies.

%---------------------------------------------------------------------
\subsection{Data Obtaining}
%---------------------------------------------------------------------

Usually, GW waveforms are divided into three stages: inspiral, merger, and
ringdown. The signal detected by a single GW detector is,
\begin{equation}
	\label{eqn:GW}
	h(t)=F^{+}(t)\,h_{+}(t)+F^{\times}(t)\,h_{\times}(t) \,,
\end{equation}
where $h_{+}$ and $h_{\times}$ are two polarization modes of GWs, $F^{+}$
and $F^{\times}$ are the corresponding pattern functions of these two
polarization modes as functions of the sky localization and the
polarization angle~\cite{Cutler:1994ys}.

In this work, we focus on GW signals generated by BBH mergers. We use the
effective-one-body numerical-relatively (EOBNR) model with aligned
spins~\cite{Bohe:2016gbl} to simulate the waveform. Without losing
generality, in addition to the SNR of GW signals, we focus on intrinsic
parameters, i.e. masses and spins. The extrinsic parameters, such as the
polarization angle and the sky localization, are all fixed to fiducial
values. The possible precession effect of BBHs is not considered, since we
are using the aligned-spin waveform. The spin parameter is denoted by
$\chi$. We set the luminosity distannce $D_{\rm L} = 100 \, {\rm Mpc}$, and
neglect the redshift effect of the GW signal. Such a setting makes us focus
on the machine learning algorithm, and the assumptions can easily be
relaxed when needed. \LS{Worth to mention that, in practice we have tested
the effects brought by the inclusion of extrinsic parameters, such as sky
location of the GW source, inclination of the BBH orbit, and the
polarization angle of the GW. Consistent optimization effects were obtained
with extrinsic parameters. Because the dependence of the GW waveform on
extrinsic parameters is much simpler than that of intrinsic parameters for
GWs from spin-aligned BBHs, in the following we will focus on the intrinsic
parameters. It is straightforward to augment with extrinsic
parameters in our machine learning data analysis.}

We use the open-source tool provided by \citet{Gebhard:2019ldz} to generate
data. This tool generates GW signals based on
PyCBC\footnote{\url{https://pycbc.org/}}~\cite{Nitz:2020:pycbc} and
LALSuite\footnote{\url{https://github.com/lscsoft/lalsuite}} platforms.
With given parameters, analog signals from LALSuite contain two time
series, i.e. the two polarization modes of GW signals. The above sequences
are combined with the corresponding antenna functions $F^{+,\times}$
according to Eq.~(\ref{eqn:GW}). The signal offset caused by the distance
difference between LIGO's Hanford and Livingston detectors is properly
introduced.

The final GW signal sequence used is,
\begin{equation}
	s(t)=h(t)+n(t) \,,
\end{equation}
where $h(t)$ is the GW waveform obtained by the above
simulation~\eqref{eqn:GW}, and $n(t)$ is the detector's noise to the
strain. The advanced LIGO's (aLIGO's) power spectral density (PSD) at the
``zero-detuned high-power'' design sensitivity
(aLIGOZeroDetHighPower)~\cite{Shoemaker:2010} is used to simulate the
Gaussian white noise. After inserting the analog waveform into the noise,
we can calculate the SNR of the strain. A rescaling of it, corresponding to
a rescaling in the distance, can achieve other desired SNR
values~\cite{Cutler:1994ys}. Finally, we get the GW strain with specific
SNR values. The strain needs to be preprocessed before being used as the
final data, which is similar to PyCBC's GW data processing. Preprocessing
stage includes two steps. The first one is data whitening. The aLIGO's
design sensitivity is used to whiten the original strain, and to filter out
the spectral components of the environmental noise so as to properly scale
its influence on the strain. The second step is filtering. We filter out
the frequency components below 20\,Hz to eliminate the influence of the
Newtonian noise in the low frequency.

\subsection{Dataset Building}
\label{subsec:building}

This work involves experiments on multiple datasets which have the same
structure. They have the following characteristics: (i) with the given
parameter range, the data parameters, masses $m_{1,2}$ and spins
$\chi_{1,2}$, are all in the form of random sampling; (ii) the ratio of the
samples containing the GW signal and pure noise in the dataset is 1 to 1;
(iii) the duration of each sample is 1\,second, and the sampling rate is
4096\,Hz, that is, each sample is a time series with a length of 4096; (iv)
considering the symmetry of mass parameters in a BBH, we use $m_1 \geq m_2$
convention to sample masses; (v) the time-series data of the samples all
use the Hanford detector sequence (the H1 sequence); (vi) in this work we
do not consider the influence of sky position of the GW signal in the
strain, that is to say, the peak value of the GW signal is located in the
same position of the strain in the datasets. These specifics are natural
for a study of such kind.

Totally, we construct 5 datasets for training and 5 datasets for testing,
which are annotated as {\it Training Datasets} and {\it Testing Datasets}
hereafter.
\begin{itemize}
\item
{\it Training Datasets}. 
Each Training Dataset is consisted of a training set and a testing set. The
training set and the testing set contain 5000 and 500 samples respectively.
When we train the model on each dataset, the model is first trained on the
training set, and then tested on the testing set. 
\item
{\it Testing Datasets}. Testing Datasets contain many sub-testing sets
(annotated as sub-datasets below) to test the model performance on
different parameter ranges. Each sub-dataset contains 500 samples.
\end{itemize}

%h-here, t-top, b-bottom，p-page-of-its-own，!-忽略“美学”标准
\def\arraystretch{1.5}
\begin{table}%[htbp!]
	\caption{Parameters for  Training Datasets.}
	\begin{tabular}{p{1.3cm}p{1.5cm}p{1.5cm}p{2.2cm}p{1.2cm}}
    \hline\hline
	Dataset & $m_1$ $(M_{\odot})$    & $m_2$ $(M_{\odot})$ 	&  $\chi_{1,2}$ & SNR         \\ 
	\hline
	1.1     & $[10,80]$ 				& $[10,80]$ 				& 0              		& $[7,7.5]$ \\
	1.2     & $[10,80]$ 				& $[10,80]$ 				& 0           			& $[7,15]$  \\
	2.1     & $[30,60]$ 				& $[30,60]$ 				& 0             		& 8           \\
	3.1     & 30          				& 45          				& $[-0.5,0.5]$ 		& 8           \\ 
	4.1     & $[10,80]$     			& $[10,80]$    			& $[-0.998,0.998]$ 	& $[7,15]$  \\
	\hline
	\end{tabular}
	\label{tab:Dataset:parameter:settings}
\end{table}

\begin{figure}%[htbp!]
		\centering
		\includegraphics[width = 7cm]{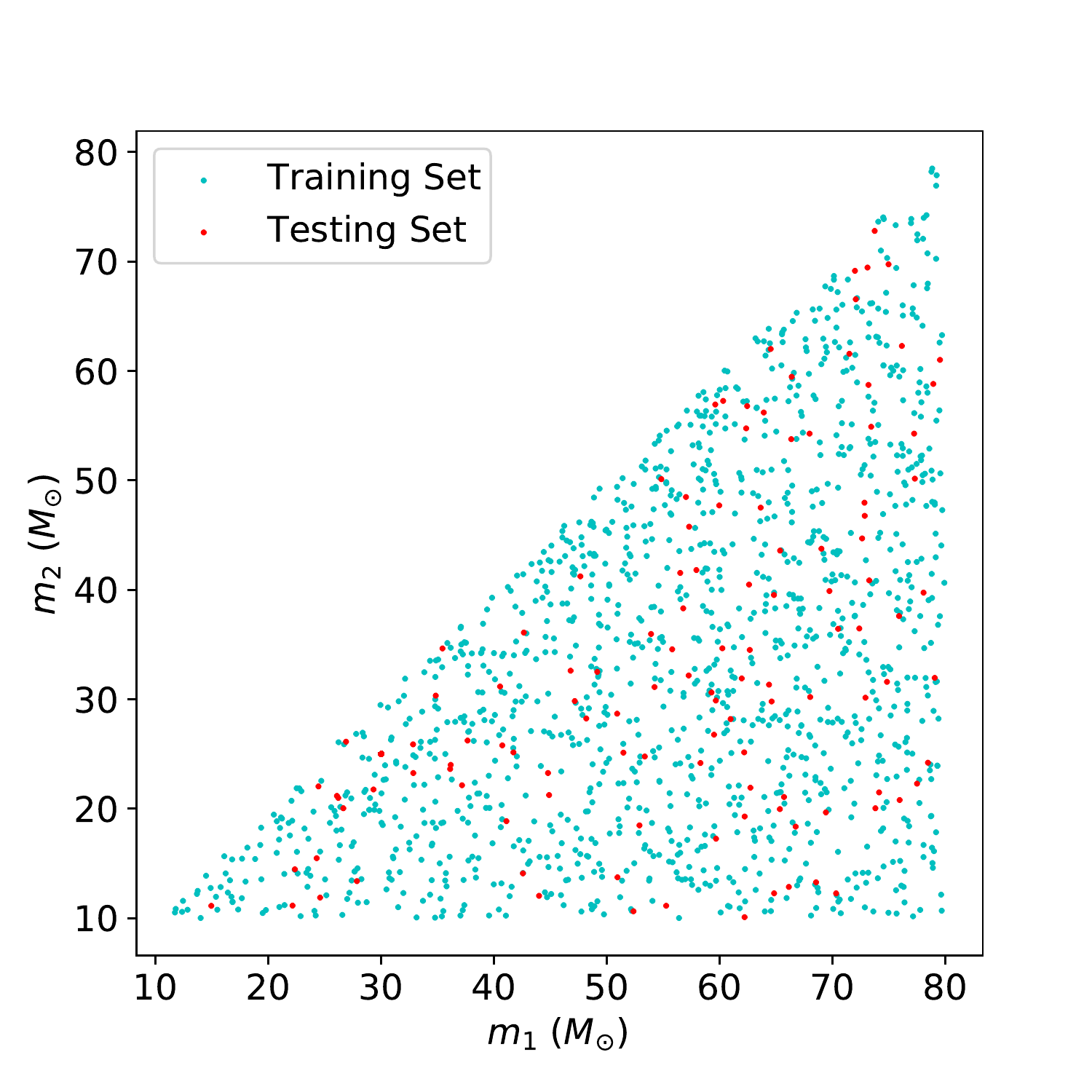}
		\caption{Mass distribution in Dataset 1.1. We use $m_1 \geq m_2$ as
		a convention to randomly sample mass parameters. Cyan points are
		the training set samples and red points are the testing set
		samples.
		\label{fig:mass:distribution:example}}
	\end{figure}

\begin{figure}%[htbp!]
		\centering
		\includegraphics[width = 8.5cm]{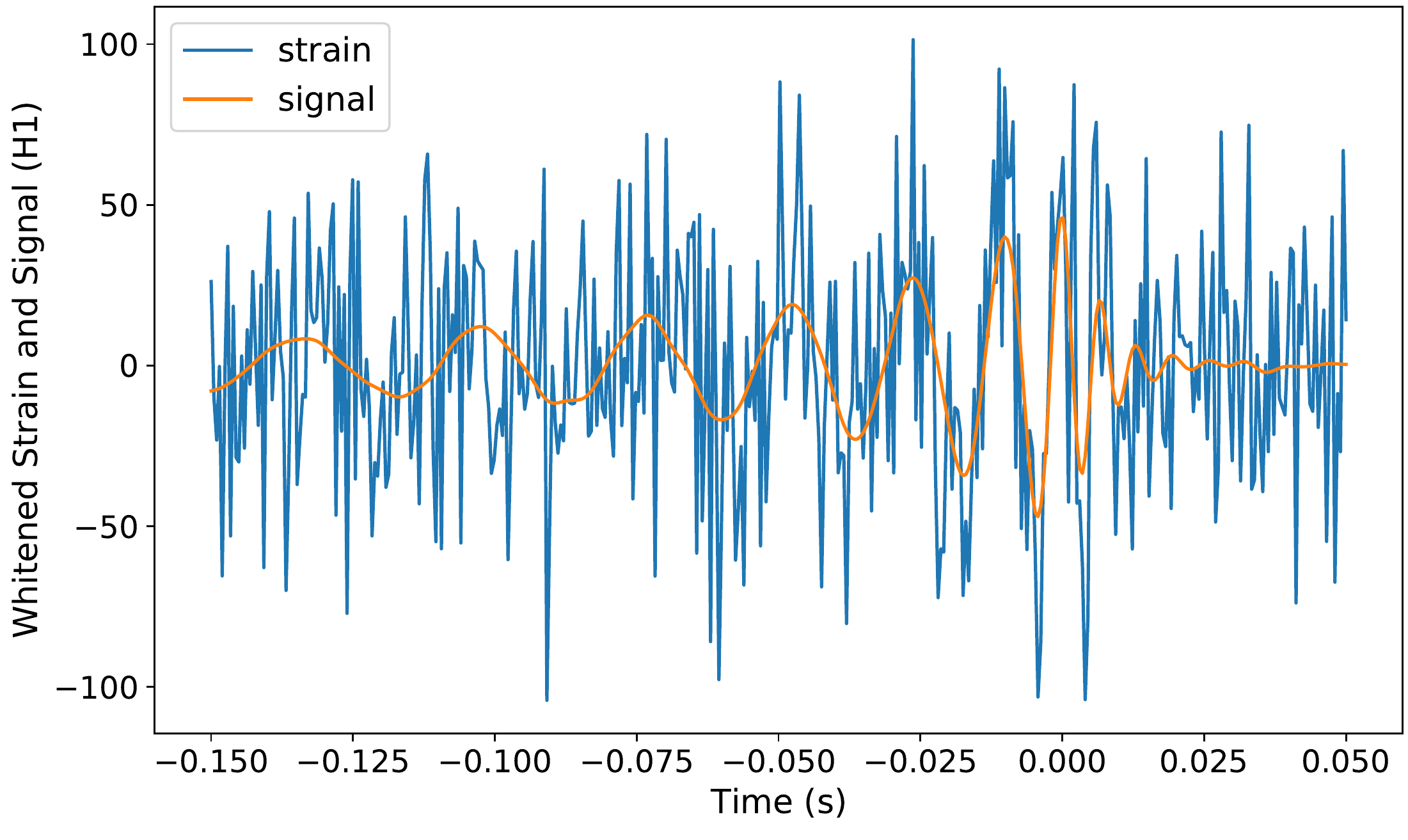}
		\includegraphics[width = 8.5cm]{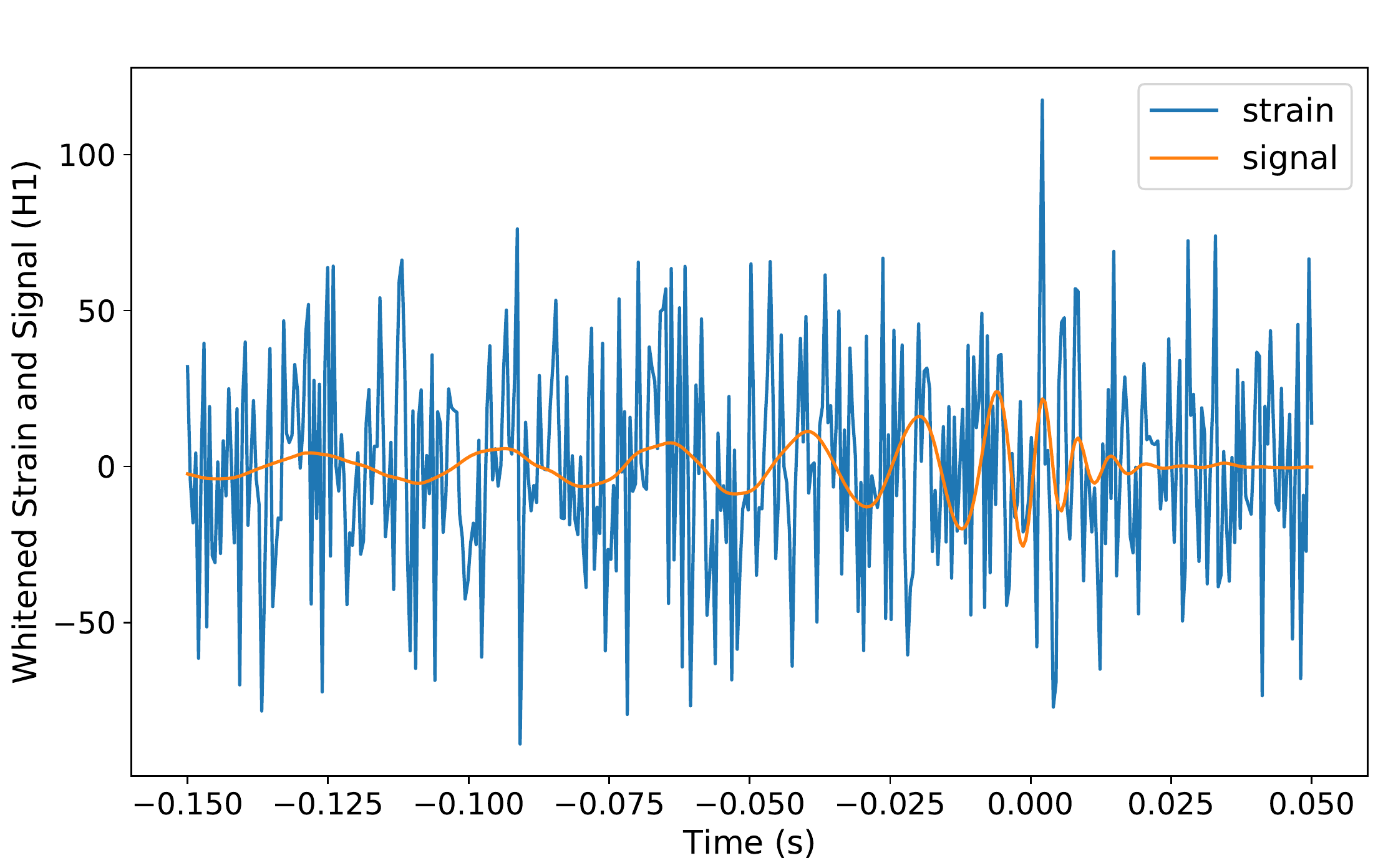}
		\caption{Examples for the simulated GW strain: $(m_1, m_2) =
		(75.61\,M_\odot, 18.64\,M_\odot)$ and $\rho=14.56$ (upper); $(m_1,
		m_2) = (67.98\,M_\odot, 27.73\,M_\odot)$ and $\rho=7.41$ (lower).
		The orange line is the normalized GW signal in the sample strain. }
		\label{fig:gravitational:wave:example}
\end{figure}

The parameters for Training Datasets used in this work are shown in
Table~\ref{tab:Dataset:parameter:settings}. Taking Dataset 1.1 as an
example. This dataset only considers three parameters: component masses of
the BBHs, $m_1$ and $m_2$, and the SNR $\rho$. The spin parameters of the
BBHs are set to zero. The mass range of both black holes is $m_i \in \left
[ 10\,M_{\odot }, 80\,M_{\odot }\right ]$ $(i=1,2)$, and the range of SNR
is $\rho \in \left [ 7,7.5\right ]$. The training set has 5000 samples in
terms of physical parameters, including 2500 samples which contain the
simulated GW signal and 2500 samples of pure noise. The testing set
contains 500 samples, including 250 samples containing the simulated GW
signal and 250 pure-noise samples. The mass parameter of each sample is
randomly drawn from a given range ($m_1 \geq m_2$ as defaulted), whose
distribution is shown in Fig.~\ref{fig:mass:distribution:example}.
Figure~\ref{fig:gravitational:wave:example} gives two GW examples from
Dataset 1.2 with a large SNR (upper panel) and a small SNR (lower panel).

\begin{table}%[htbp!]
	\caption{Parameters for Testing Datasets.}
	\label{tab:Testing:dataset:parameters:setting}
	\begin{tabular}{p{1cm}p{1.5cm}p{1.5cm}p{2.7cm}p{1.3cm}}
	\hline\hline
	Dataset & $m_1$ $(M_{\odot})$    & $m_2$ $(M_{\odot})$ 	& $\chi_{1,2}$  & SNR         \\
	\hline
	1               & $[10,80]$    		    & $[10,80]$    		   & 0                       & $[7,0.5,15]$ \\
	2               & $[10,10,80]$ 			& $[10,10,80]$ 		   & 0                       & 8              \\
	3               & 30            			& 45             		   & $[-0.998,0.25,0.998]$ & 8              \\
	4               & $[10,80]$   			& $[10,80]$    		   & $[-0.998,0.998]$      & $[7,15]$ \\ 
	5               & $[10,80]$   			& $[10,80]$    		   & $[-0.998,0.998]$      & $[7,0.5,15]$ \\
	\hline
	\end{tabular}
\end{table}

The parameter settings of Testing Datasets are shown in
Table~\ref{tab:Testing:dataset:parameters:setting}. As shown above, the
parameters used to construct each sub-dataset are given in the $[{\tt min},
{\tt step}, {\tt max}]$ format, where ${\tt min}$ and ${\tt max}$ define
the overall sampling range of the parameter, and each sub-dataset is
constructed with a uniform ${\it step}$ size. Take the Testing Dataset 1 as
an example, which consists of 16 sub-datasets. The SNR parameter sampling
range of the first sub-dataset is $[7,7.5]$. Each parameter is randomly
sampled within this range, and the number of samples is 500. The Testing
Dataset will be used to investigate the detection ability of the model in
different parameter intervals and explore model robustness with the change
of parameter values.

%---------------------------------------------------------------------
\section{Improving Techniques}
\label{sec:techniques}

\subsection{Model Baseline}
\label{subsec:baseline}

\begin{figure}%[htbp!]
		\centering
		\includegraphics[width = 7cm]{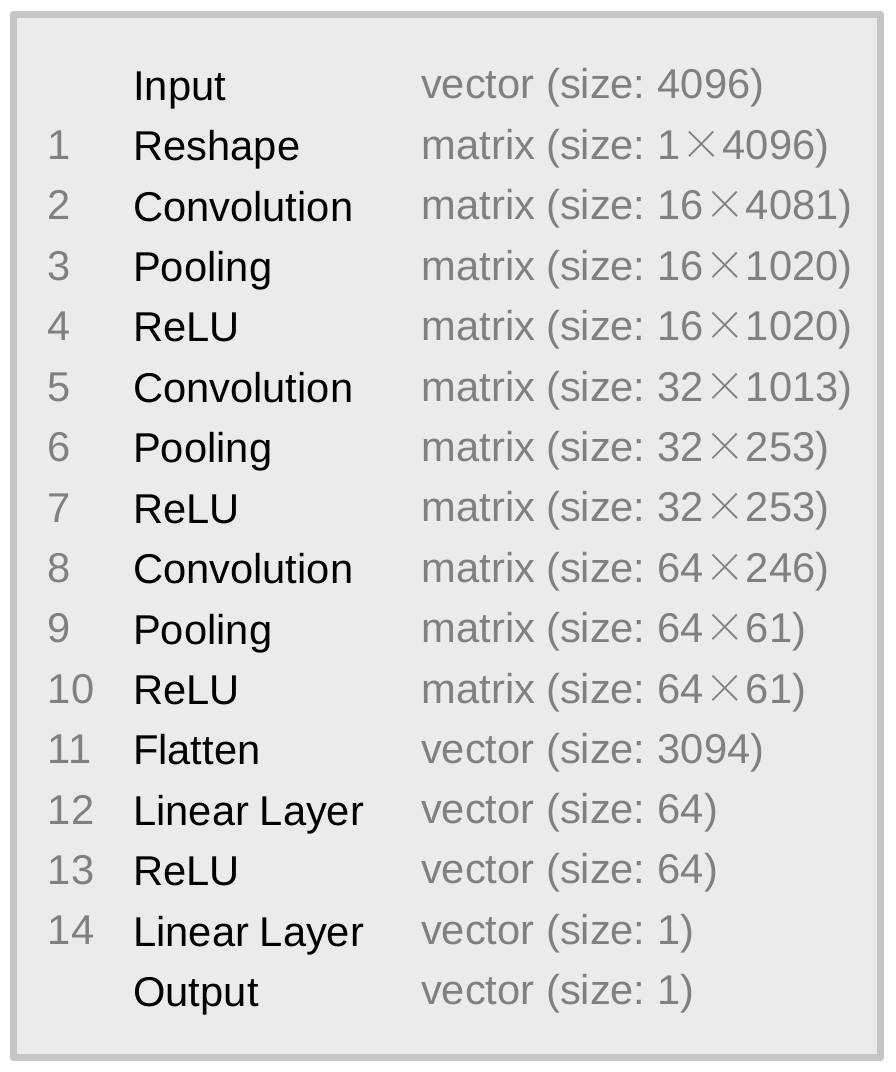}
		\caption{Basic CNN model architecture that is used in this
		work~\cite{George:2016hay}.}
		\label{fig:basic:CNN:model:architecture}
	\end{figure}

After preliminary experiments with parameters on networks with different
depths and hyperparameter values, we decide to use the network structure
similar to that of \citet{George:2016hay} as the baseline CNN model. Its
structure is shown in Fig.~\ref{fig:basic:CNN:model:architecture}. 

The model receives GW strain as input and outputs the discriminant value of
classification. In the model, the classification threshold is set to 0.5,
that is, a sample with a discriminant value greater than 0.5 is judged as a
positive sample (i.e., containing a GW signal), otherwise it is judged as a
negative one (i.e., pure noise). The baseline model contains three
convolutional layers. The number of channels of each convolutional layer is
16, 32, and 64, and the size of the convolution kernel for them is 16, 8,
and 8, respectively. After each convolutional layer, a pooling layer and an
activation layer are provided. The pooling layer uses maximum pooling with
a pooling window of 4 and a step size of 4, which means that the feature
sequence is down-sampled to a quarter of the original sequence length and
retains the feature with the largest value. The active layer uses the {\tt
ReLU} function shown in Eq.~(\ref{eq:ReLU}). Subsequently, the feature
sequences extracted by the convolution structure are input to the fully
connected layer to achieve discriminative classification. The model outputs
discriminant values to determine whether the sequence contains a GW signal.

In this work, cross entropy is used as the loss function to update the
gradient. Cross entropy is an important concept proposed by
\citet{Shannon:1948} in information theory. It is often used to measure the
difference between the predicted distribution and the true
distribution~\cite{Shannon:1948}. Let the label of each sample be $y$ and
the discriminant value given by the model be $p$. The cross entropy is,
\begin{equation}
	J=-\big( y\cdot \log p+(1-y)\cdot \log(1-p) \big) \,.
\end{equation}

We use {\tt ADAM}~\cite{Kingma:2014} as a gradient descent strategy to
update our model parameters. This strategy inputs the samples into the
model in batches to calculate the gradient and updates the parameters.
After trying batch number values of 5, 10, 25, 50, 100, 200, 250, and 500,
we take the value 25, on which the model has the highest accuracy. The
learning rate is set to $5\times10^{-5}$, and the training rate is reduced
10 times every 20 epochs to avoid overfitting the model.

The code implementation of our work is based on the {\tt Pytorch}
framework~\cite{Paszke:2019}, which uses the {\tt CUDA} deep learning
library (cuDNN)~\cite{Chetlur:2014} to accelerate the GPU's model
operation. Our work deploys experiments on NVIDIA TITAN X GPU.

\subsection{Improving Techniques}

We now discuss several improvements---dropout, batch normalization, and the
$1\times1$ convolution---that we experiment on the model of
\citet{George:2016hay}.

% \subsubsection{Dropout}

{\bf Dropout} was first proposed by ~\citet{Hinton:2012} in 2012 to improve
the overfitting problem of neural networks. Subsequently, dropout has
become one of the widely used techniques in deep
learning~\cite{Srivastava:2014}. Its basic idea is that, in each batch of
training, the probability $p$ is artificially specified, so that the
neurons in the fully connected layer stop working with the probability $p$,
and set their parameter values are set to zero.

The advantages of dropout are the following~\cite{Hinton:2012}. First, in
the training process of each batch, because the neurons of each layer are
inactivated with probability $p$, the network structure of each training is
different. The overall model training is equivalent to the joint
decision-making process between multiple neural networks with different
structures, which helps to improve the problem of overfitting. Second,
dropout results in that, two neurons do not necessarily appear in the same
network structure each time so that the parameter update no longer depends
on the joint decision of some neurons with fixed relationships. At the
feature level, this technique prevents decision-making from over-dependence
on certain features and forces the model to learn more robust feature
representations~\cite{Goodfellow:2016}.

% \subsubsection{Batch Normalization}

{\bf Batch normalization} is a neural network training technique proposed
by~\citet{Ioffe:2015} in 2015. Its specific idea is the following. In the
training process of each batch, after the data passes through the
activation layer, the activation value of each batch of data is normalized.
That is, the average value of the sample data of each batch is normalized
to 0, and the variance is normalized to 1. In a batch of data of length
$m$, the activated data are set to $\mathcal{B}=\left\{ x_1, \cdots, x_m
\right \}$, and the batch normalized data are set to $\boldsymbol y = \left
\{ y_1, \cdots , y_m \right \}$. Then the batch normalized algorithm is
expressed as~\cite{Ioffe:2015}
%--
\begin{align}
\mu_{\mathcal{B}} &\leftarrow \frac{1}{m} \sum_{i=1}^{m} x_{i} \,, \\
\sigma_{\mathcal{B}}^{2} &\leftarrow \frac{1}{m} \sum_{i=1}^{m}\left(x_{i}-\mu_{\mathcal{B}}\right)^{2} \,,\\
\widehat{x}_{i} &\leftarrow \frac{x_{i}-\mu_{\mathcal{B}}}{\sqrt{\sigma_{\mathcal{B}}^{2}+\epsilon}} \label{eq:hatx} \,,\\
y_{i} &\leftarrow \gamma \widehat{x}_{i}+\beta \,,
\end{align}
%--
where $\gamma$ and $\beta$ in the algorithm are the parameters learned in
the gradient update. The purpose of this step is to make the result of
batch normalization be the same as the original input data, which maintains
a possibility to retain the original structure. The parameter $\epsilon$ in
Eq.~(\ref{eq:hatx}) is used to prevent invalid calculation when the
variance $\sigma_{\mathcal{B}}^{2}$ is zero.

The batch normalization technique makes the mean and variance of the input
data distribution of each layer in the CNN within a certain range. Thus,
each layer of the network does not need to adapt to the change in the
distribution of input data, which is conducive to accelerating the learning
speed of the model and speeding up the simulation~\cite{Goodfellow:2016}.
At the same time, batch normalization suppresses the problem that small
changes in parameters are amplified with the deepening of the network
layers, making the network more adaptable to model parameters and more
stable gradient updates. Besides, due to the use of dropout technique, the
number of effective neurons in the model decreases, and the fitting speed
of network slows down~\cite{Ioffe:2015}. Considering that batch
normalization has a significant improvement effect on the fitting speed of
the model, dropout and batch normalization techniques are often introduced
into the structure of the neural network at the same time~\cite{li:2019}.

% \subsubsection{$1\times1$ Convolution}

{\bf The $1\times1$ convolution}, also known as ``network in network'', was
proposed by \citet{Lin:2013} in 2014. The $1\times1$ convolution is a
convolution kernel of size $1\times1$. In one-dimensional convolution, it
is a convolution kernel in which the window length is 1. The $1\times1$
convolution is similar to the traditional convolutional layer. After
receiving the input data, the operation is gradually performed to obtain
the output feature map representation. Its setting parameters commonly used
are the number of input channels and the number of output
channels~\cite{Paszke:2019}. The number of input channels is the feature
dimension of the original data. The number of output channels is the same
as the number of convolution kernels, which is the feature dimension of the
output data.

The $1\times1$ convolution recombines multi-dimensional features of the
original data without changing the length of them. It enhances the model's
ability to express data features~\cite{Lin:2013}. In addition, if the
number of input channels is fewer than the number of output channels, then
the $1\times1$ convolution can be seen as an ascending dimension operation
of the original data to increase the feature dimension. This is similar to
the effect of data passing through a fully connected layer. Through
$1\times1$ convolution, the interaction and reorganization of
multi-dimensional features of the original data are realized, and the
model's ability to represent data features is
enhanced~\cite{Goodfellow:2016}.

%---------------------------------------------------------------------
\section{Simulation Results}
\label{sec:res}
%---------------------------------------------------------------------

\begin{table}%[htbp!]
	\caption{The extended techniques used for each model.}
	\label{tab:model:architecture}
	\begin{tabular}{p{2cm}p{2cm}p{2cm}p{2cm}}
    \hline\hline
	Model    & Dropout & Batch norm. & $1\times1$ conv. \\ 
	\hline
	ConvNet1 &         &                     &          \\
	ConvNet2 & $\surd$ &                     &          \\
	ConvNet3 &         & $\surd$             &          \\
	ConvNet4 &         &                     & $\surd$  \\
	ConvNet5 & $\surd$ & $\surd$             &          \\
	ConvNet6 & $\surd$ & $\surd$             & $\surd$  \\
	\hline
	\end{tabular}
\end{table}

\begin{figure*}%[htbp!]
	\centering
	\includegraphics[height = 5.6cm]{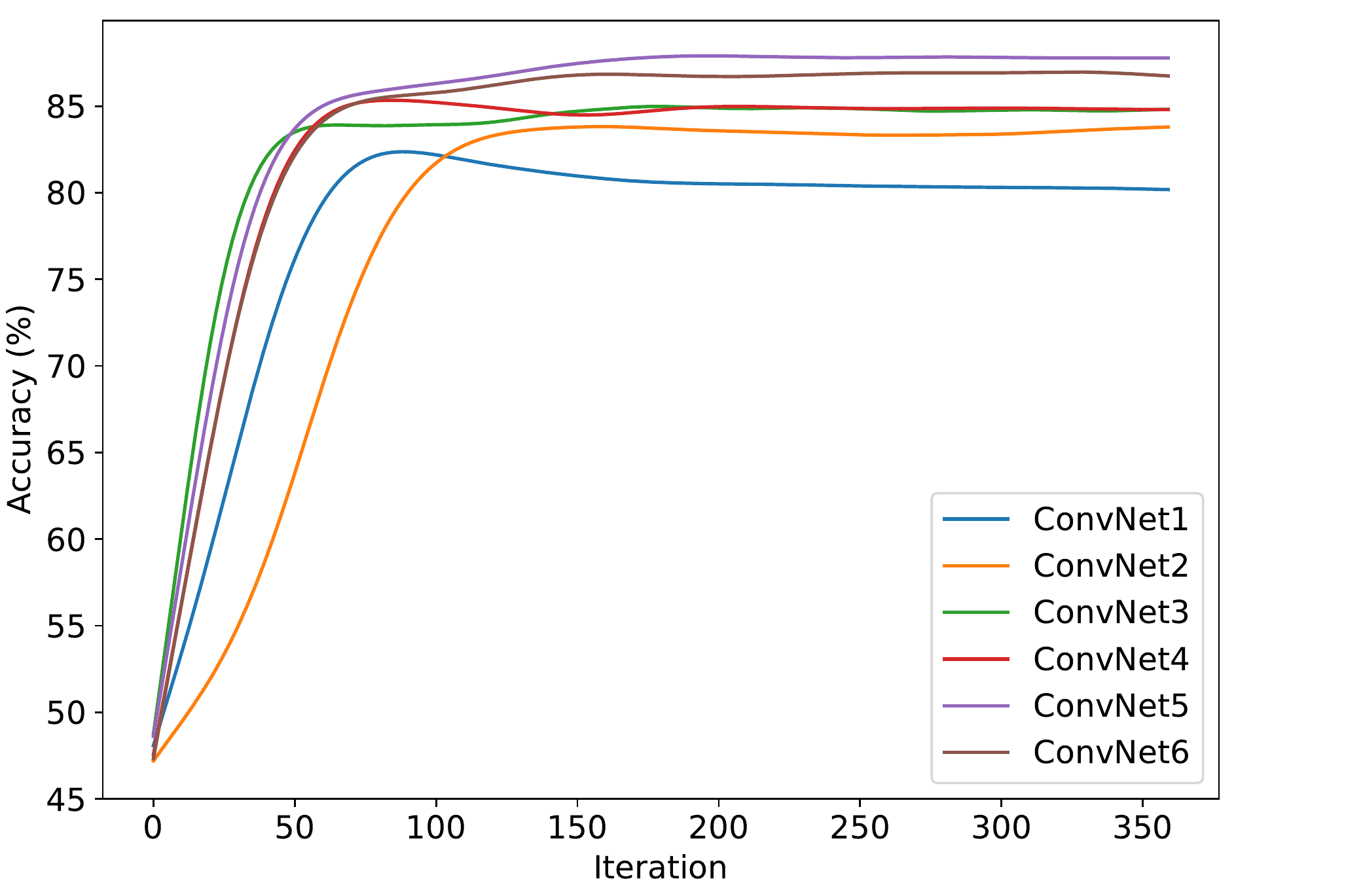}
	\includegraphics[height = 5.6cm]{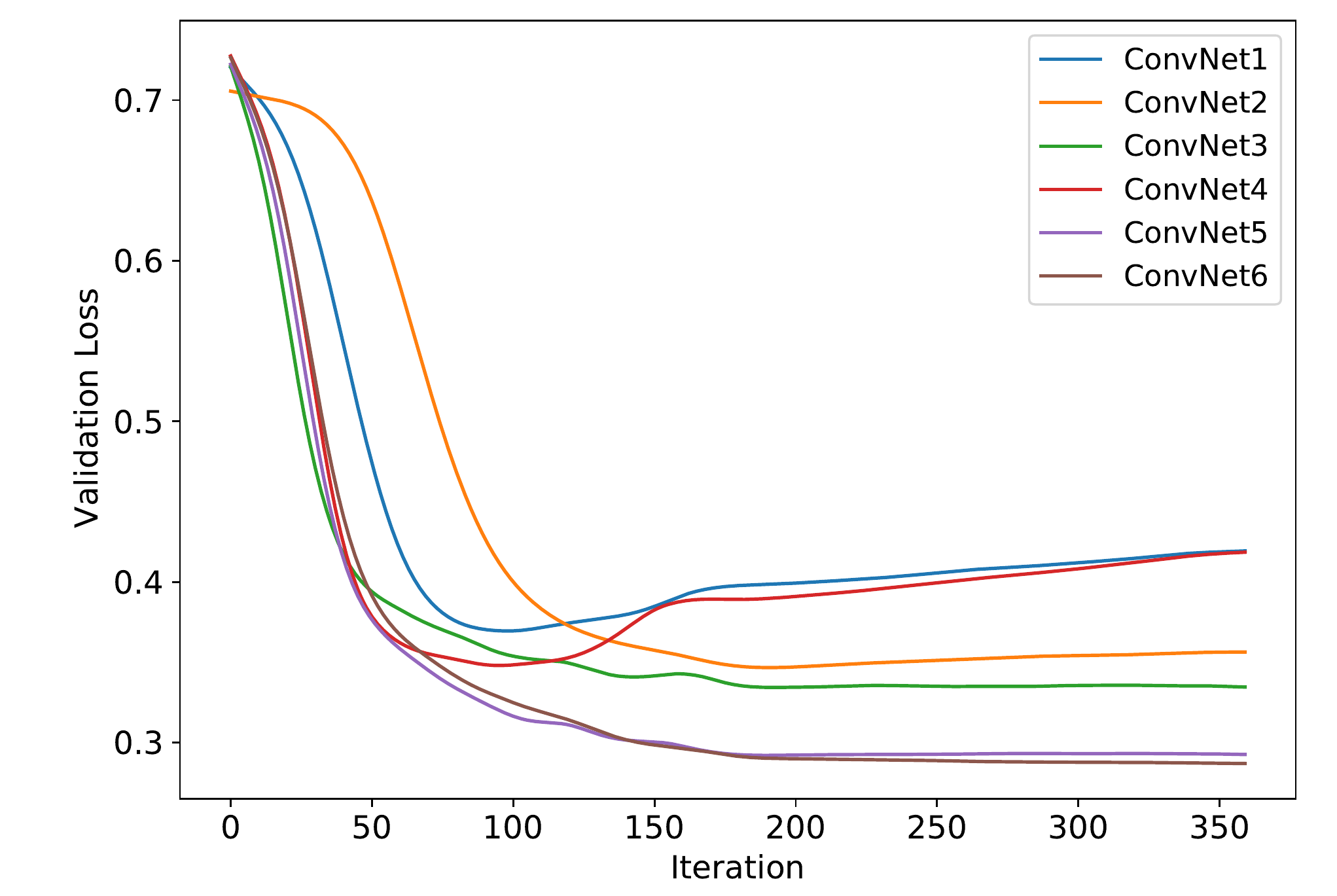}
	\includegraphics[height = 5.6cm]{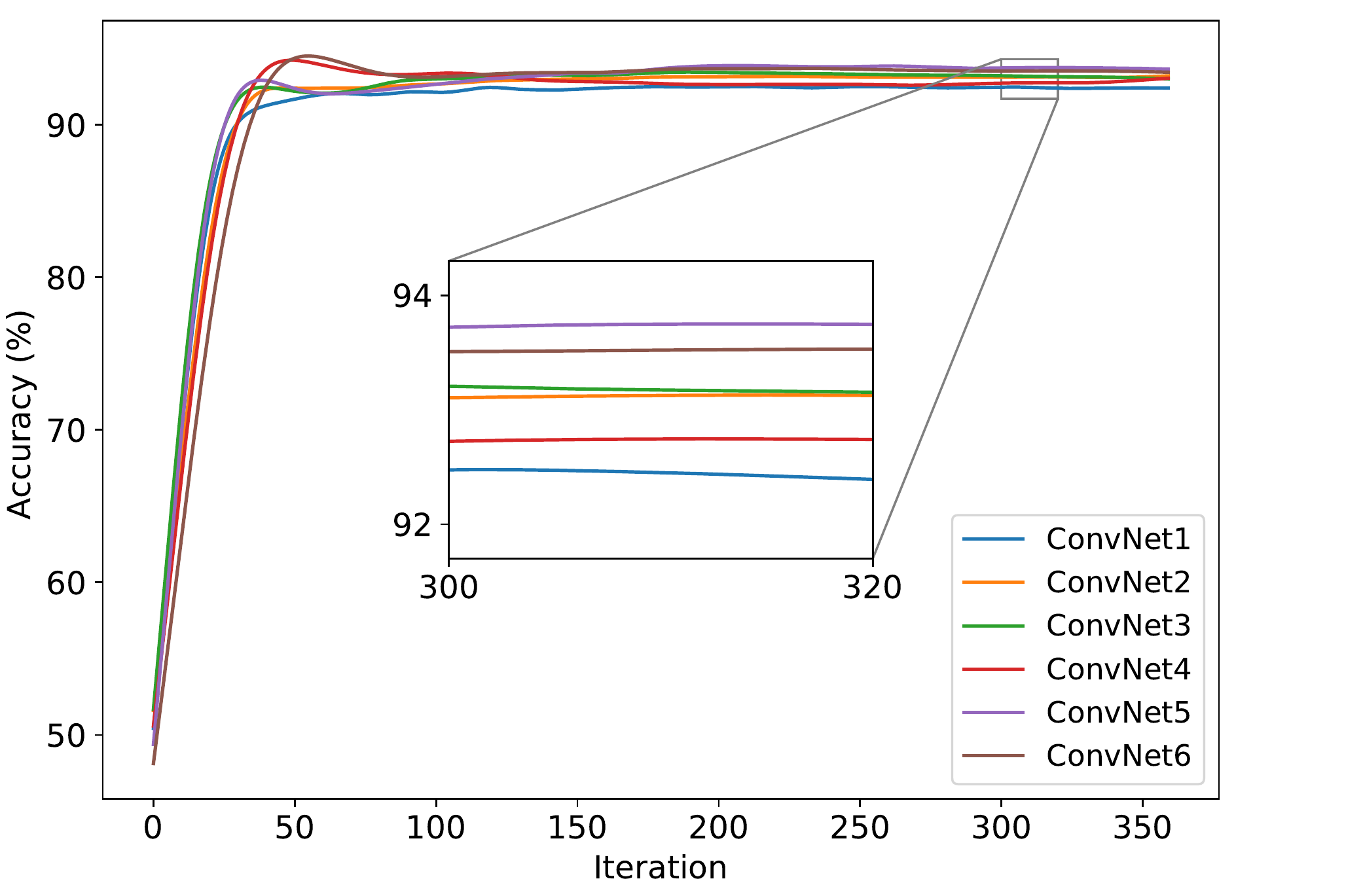}
	\includegraphics[height = 5.6cm]{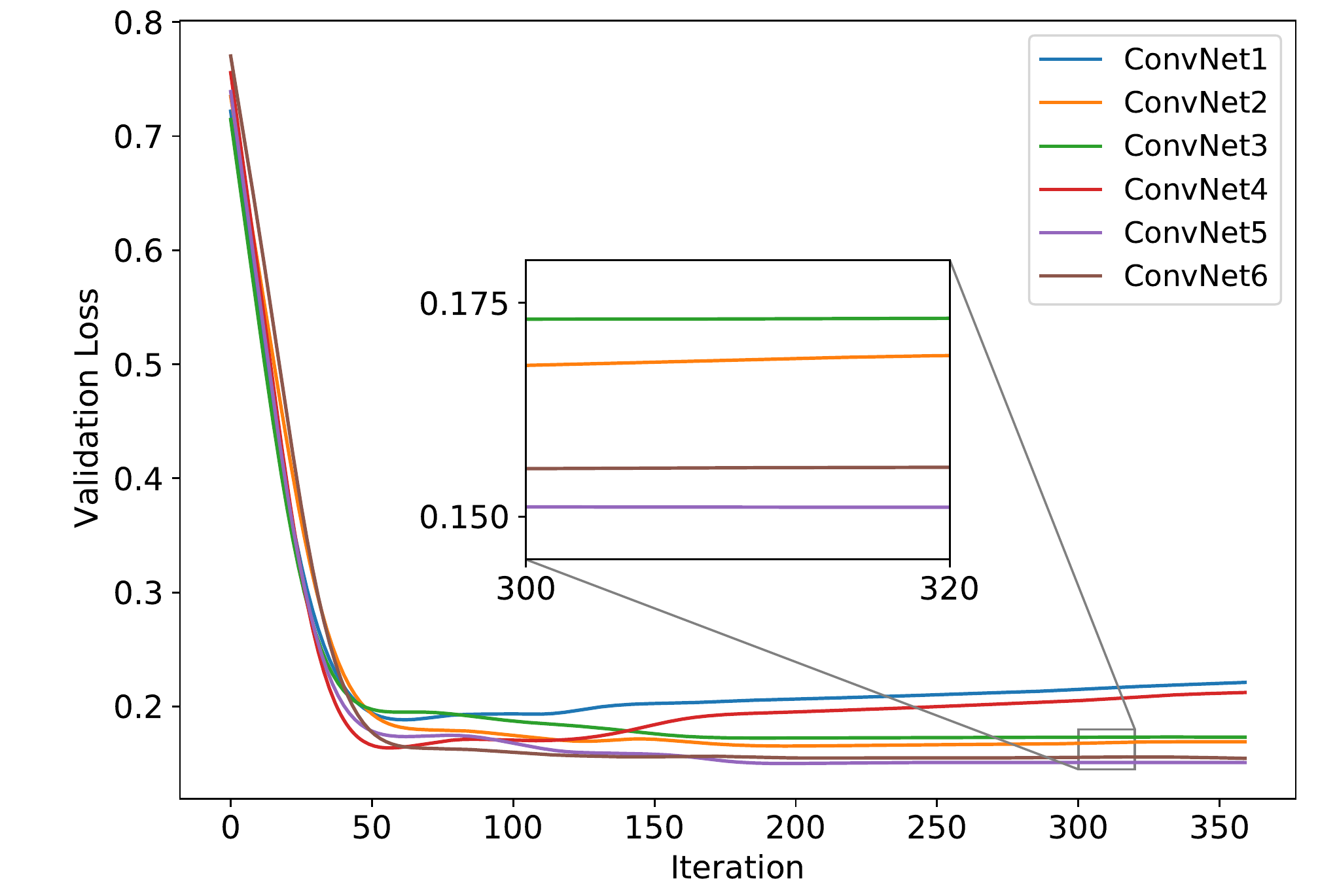}
	\caption{Model comparison on Datasets 1.1 (upper panels) and 1.2 (lower
	panels). ConvNet1 to ConvNet6 are the models shown in
	Table~\ref{tab:model:architecture}. We use the accuracy (fraction of
	samples correctly classified) and the validation loss (loss function
	value in the validation set) as our metrics to track the model
	performance in the training process.}
	\label{fig:model:comparison:on:Dataset1p1}
\end{figure*}

Based on the enhanced techniques, we extend the basic model of CNN
described in Sec.~\ref{subsec:baseline}. Dropout, batch normalization, and
the $1\times1$ convolution are successively added to the basic model. We
test these models with Datasets 1.1 and 1.2. In the Dataset 1.1, the SNR
range of GW signals is $\rho \in [7,7.5]$, which makes the signal hard to
be detected, representing the edge cases for detection on low SNR signals.
The SNR range of Dataset 1.2 is $\rho \in [7,15]$, reflecting the detection
capability of the model with relatively loud events. The extended models
are shown in Table~\ref{tab:model:architecture}. Considering that dropout
and batch normalization have complementary performance in the overfitting
problem, we combine these two techniques in our multi-technique models.

The training set and the testing set of the model are based on Datasets 1.1
and 1.2 of Training Datasets described in Sec.~\ref{subsec:building}. Model
parameters and learning strategies are consistent with
Sec.~\ref{sec:techniques}. We divide 500 samples from the training set as
the validation set, which is used to verify the model effect during the
training process. When the validation set is input into the model, it does
not participate in the gradient update, but is only used to calculate the
accuracy and loss function value.

\begin{figure*}%[htbp!]
	\centering
	\includegraphics[height = 5.3cm]{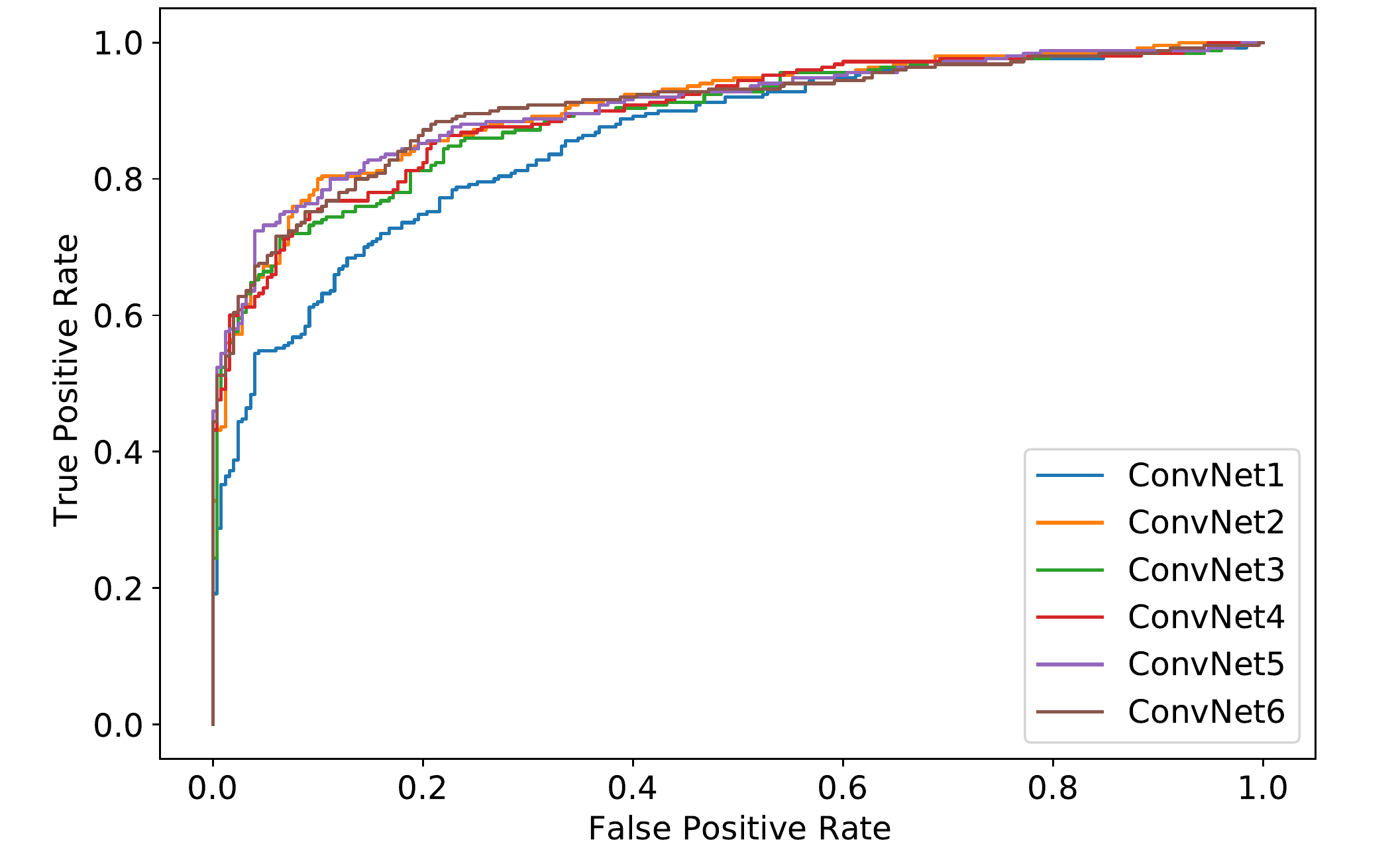}
	\includegraphics[height = 5.3cm]{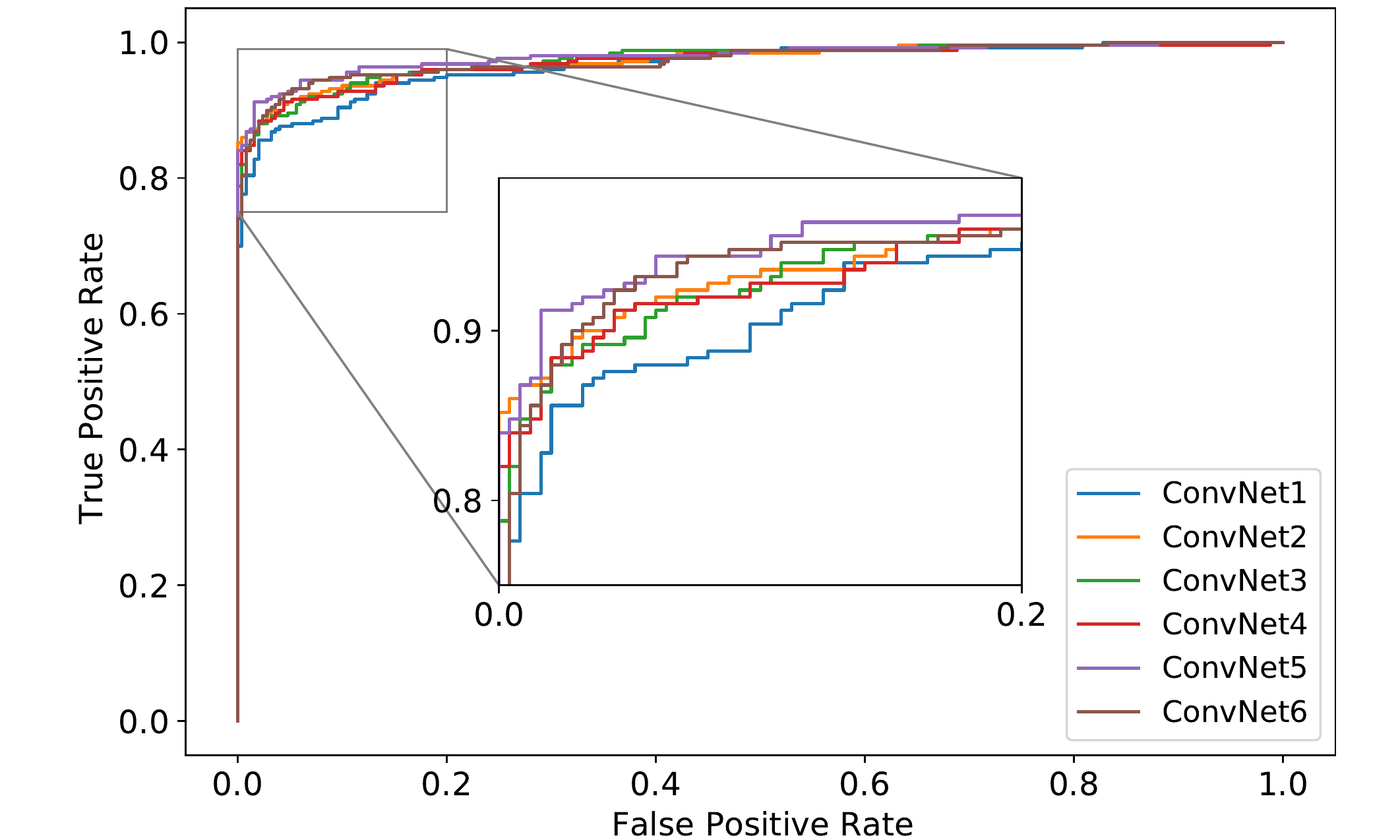}
	\includegraphics[height = 5.2cm]{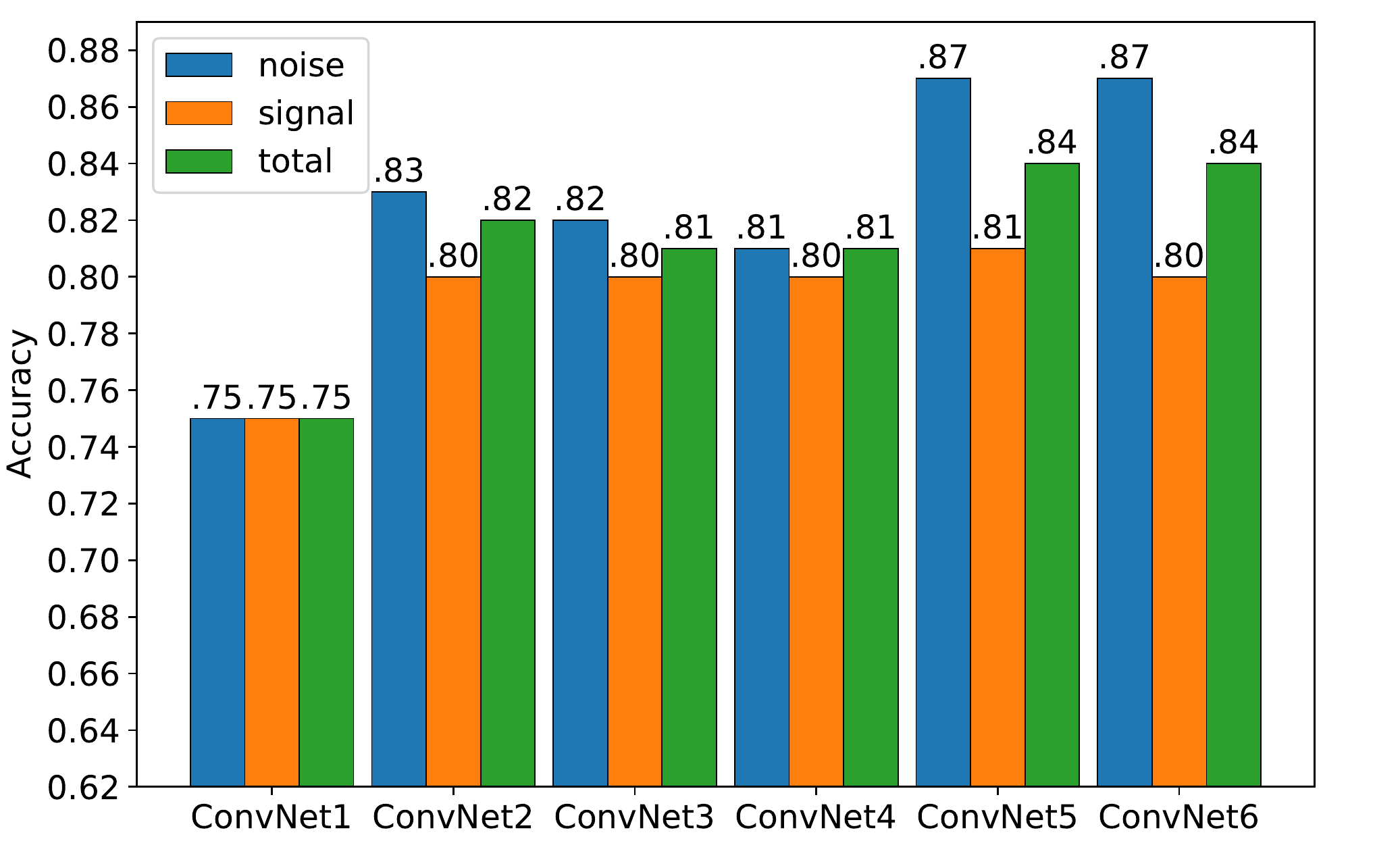}
	\includegraphics[height = 5.2cm]{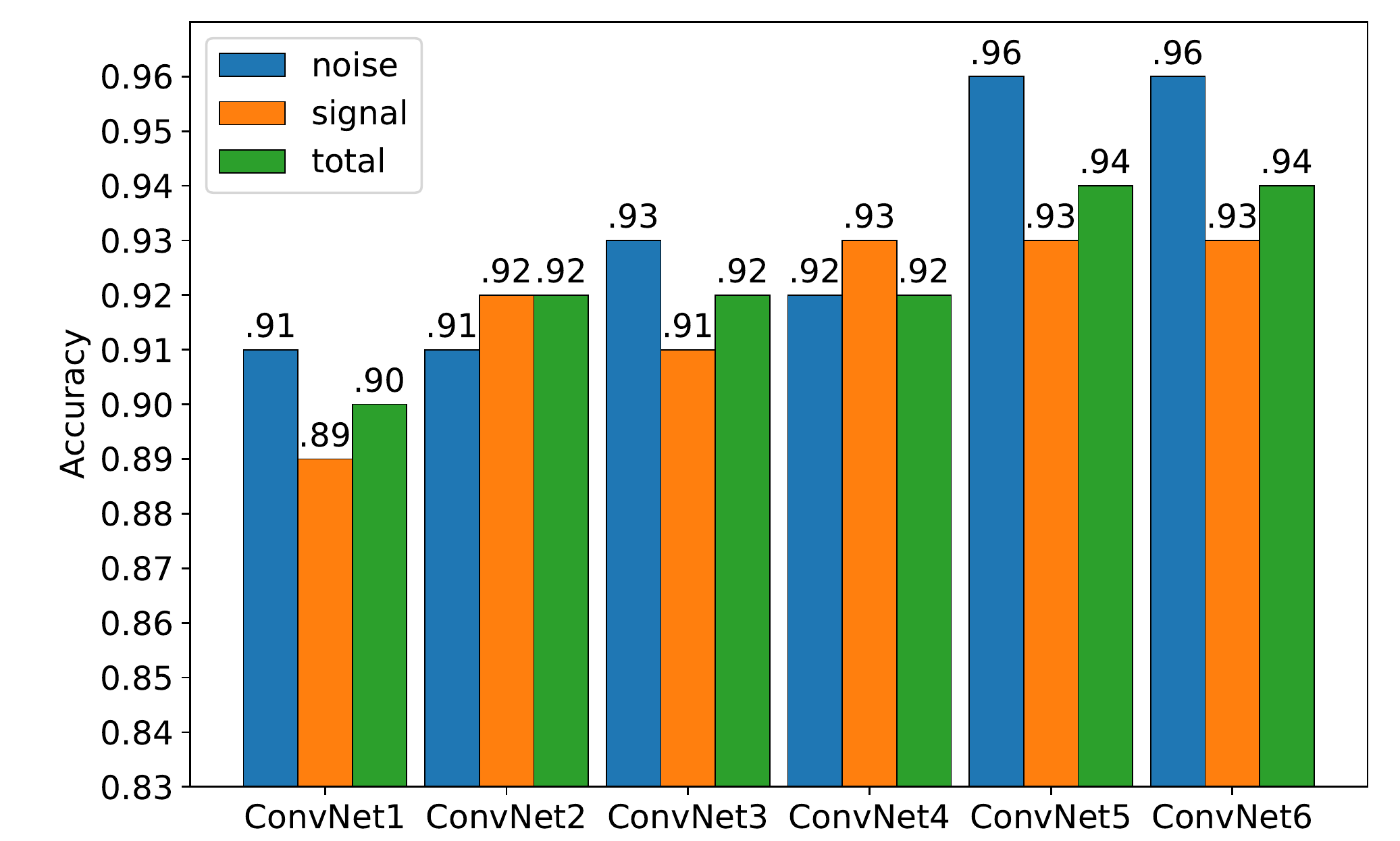}
	\caption{Results on the testing sets of Datasets 1.1 (left panels) and
	1.2 (right panels). We use both ROC and the accuracy as our evaluation
	metrics. As seen in lower panels, the accuracy in identifying noise and
	signal, as well as the total accuracy, are illustrated separately.}
	\label{fig:final:test:results:on:Dataset}
\end{figure*}

% \begin{table}%[htbp!]
% 	\caption{AUC scores on Dataset 1.1 and 1.2.}
% 	\label{tab:model:auc}
% 	\begin{tabular}{p{1cm}p{1cm}p{1cm}p{1cm}p{1cm}p{1cm}p{1cm}}
%     \hline\hline
%     Dataset  & Conv.1 & Conv.2 & Conv.3 & Conv.4 & Conv.5 & Conv.6 \\
% 	\hline
%     1.1      & 0.86   & 0.90   & 0.89   & 0.90   & \textbf{0.91} & \textbf{0.91} \\
%     1.2      & 0.96   & 0.97   & 0.97   & 0.97   & \textbf{0.98} & 0.97\\
% 	\hline
% 	\end{tabular}
% \end{table}

\begin{table}%[htbp!]
	\caption{AUC scores on Dataset 1.1 and 1.2. The highest value in each
	column is highlighted.}
	\label{tab:model:auc}
	\begin{tabular}{p{3cm}p{2cm}p{2cm}}
    \hline\hline
	Dataset  & 1.1 & 1.2 \\
	\hline
	ConvNet1 & 0.86 & 0.96 \\
	ConvNet2 & 0.90 & 0.97 \\
	ConvNet3 & 0.89 & 0.97 \\
	ConvNet4 & 0.90 & 0.97 \\
	ConvNet5 & {\bf 0.91} & {\bf 0.98} \\
	ConvNet6 & {\bf 0.91} & 0.97 \\
	\hline
	\end{tabular}
\end{table}

The validation results of each model in the training process are shown in
Fig.~\ref{fig:model:comparison:on:Dataset1p1}. As shown in the figure,
based on Dataset 1.1 (low SNR), the accuracy of each extended network model
is significantly improved compared with that of the basic model (ConvNet1).
The validation results show that the final accuracy of each network is
stable at $83\%$--$87\%$, while the accuracy of the basic model is below
$80\%$. Among them, ConvNet5, which uses dropout and batch normalization,
achieves the highest accuracy in the stable stage after enough iterations.
ConvNet6, which uses all improving techniques---namely dropout, batch
normalization, and $1\times1$ convolution---has an accuracy slightly lower
than ConvNet5. ConvNet4, which only uses the $1\times1$ convolution,
achieves a high accuracy similar to ConvNet6 in the early stage of
training, but decreases later on.

Observing the change in the value of the loss function of the validation
set, we find that, in the stable stage, ConvNet6 reaches the lowest value
in loss, and the loss of ConvNet5 is slightly higher than ConvNet6.
Similarly, the minimum values of the loss function of extended models are
all smaller than that of the basic model (ConvNet1). Specifically, after
applying the dropout to the model, the decline of the loss function
values of ConvNet2 is significantly later than the basic model, and in
the stable stage, the validation loss of ConvNet2 is lower than that of
the basic model. This shows that the dropout technique indeed slows down
the fitting speed
of network and reduces the overfitting problem. The loss function value of
ConvNet3 with batch normalization decreases faster. This means that the
fitting speed of the model is accelerated. At the same time, its loss in
validation in the late training stage is relatively stable. ConvNet4 with
the $1\times1$ convolution has a faster fitting speed among the extended
models with a single technique, but its late-stage model performance is
unstable. In addition, it has the overfitting problem similar to the basic
model. ConvNet5 with dropout and batch normalization, and ConvNet6 which
uses all three techniques, perform similarly. They both achieve the optimal
results in the stable stage. ConvNet6 performs slightly better than
ConvNet5 in the stable stage of training.
	
In Dataset 1.2, the SNR of the GW signal is larger than that in Dataset
1.1, which makes the signal detection much easier and reduces the
difference between models. In this situation, the extended model is still
superior to the basic model. In terms of accuracy, ConvNet5 reaches the
optimal value in the stable stage of the late training epoch. Its
performance is slightly better than that of ConvNet6. In terms of the value
of loss function, ConvNet5 and ConvNet6 have similar results in the stable
stage. Actually, ConvNet5 is slightly better than ConvNet6 in the stable
stage. In addition, as the SNR distribution range of the dataset expands,
the differences between the samples increase so that the overfitting
problem of the model is reduced.

After training, the models are provided to the testing sets for final
evaluation. The testing results are shown in
Fig.~\ref{fig:final:test:results:on:Dataset}. ROC curve is a graphical plot
that illustrates the diagnostic ability of a binary classifier system as
its discrimination threshold is varied~\cite{Fawcett:2006}; see
Sec.~\ref{subsec:baseline}. When the threshold changes, the curve reflects
the fraction of positive samples correctly identified (True Positive Rate)
versus the fraction of negative samples incorrectly identified (False
Positive Rate)~\cite{Powers:2008}. The area under the curve (annotated as
AUC) is equal to the probability that a model ranks a randomly chosen
positive sample higher than a negative one~\cite{Fawcett:2006}.

The AUC scores of different models on Datasets 1.1 and 1.2 are shown in
Table \ref{tab:model:auc}. It can be seen that on Dataset 1.1, ConvNet5 ans
ConvNet6 both achieve the highest accuracy and AUC score, but the
classification rate of GW signals of ConvNet6 is slightly lower than that
of ConvNet5 (shown in Fig.~\ref{fig:final:test:results:on:Dataset}). On
Dataset 1.2, ConvNet5 achieves both the highest accuracy and AUC score.

\begin{figure*}%[htbp!]
	\centering
	\includegraphics[width = 7.0cm]{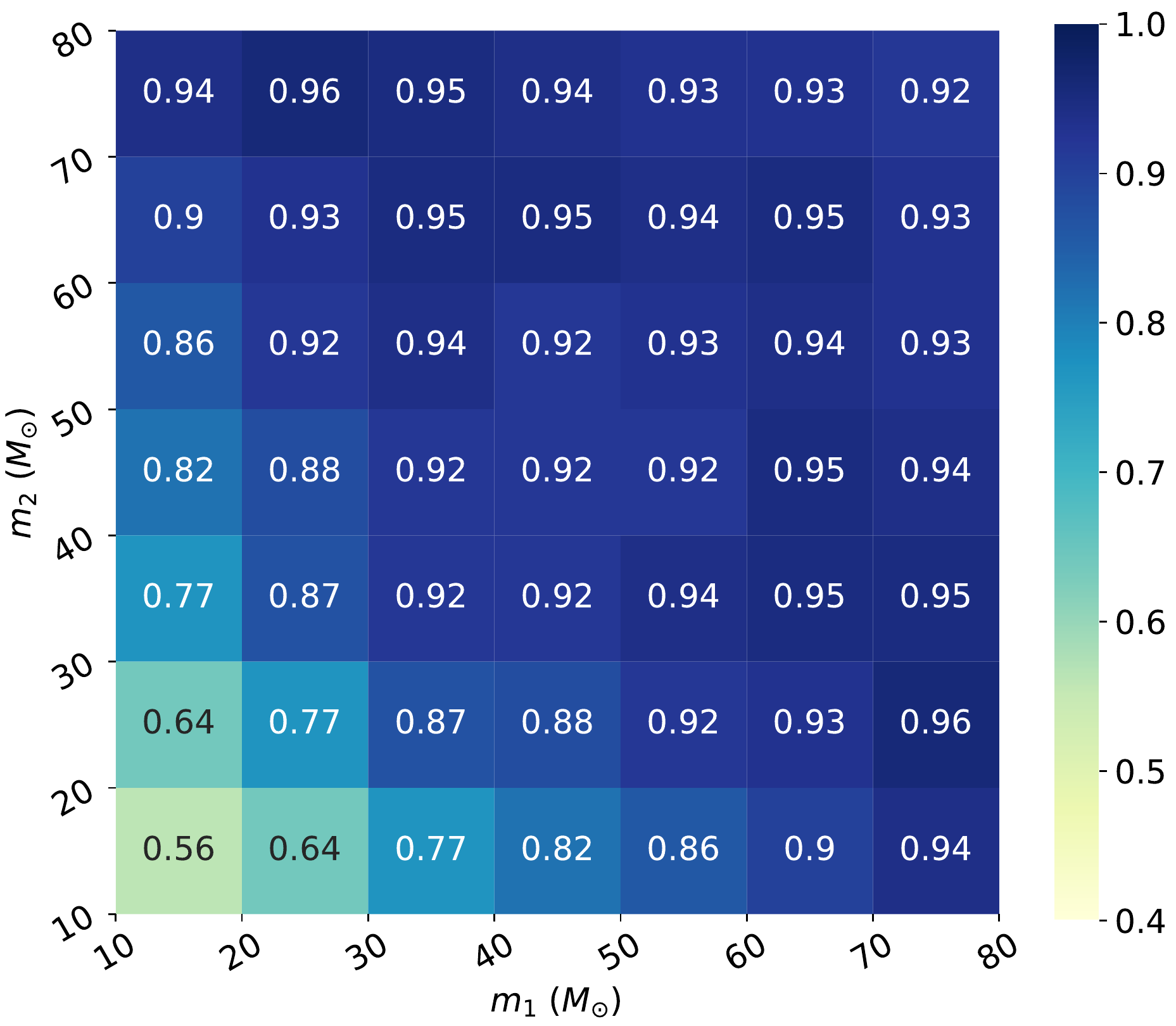}
	\hspace{18mm}
	\includegraphics[width = 7.28cm]{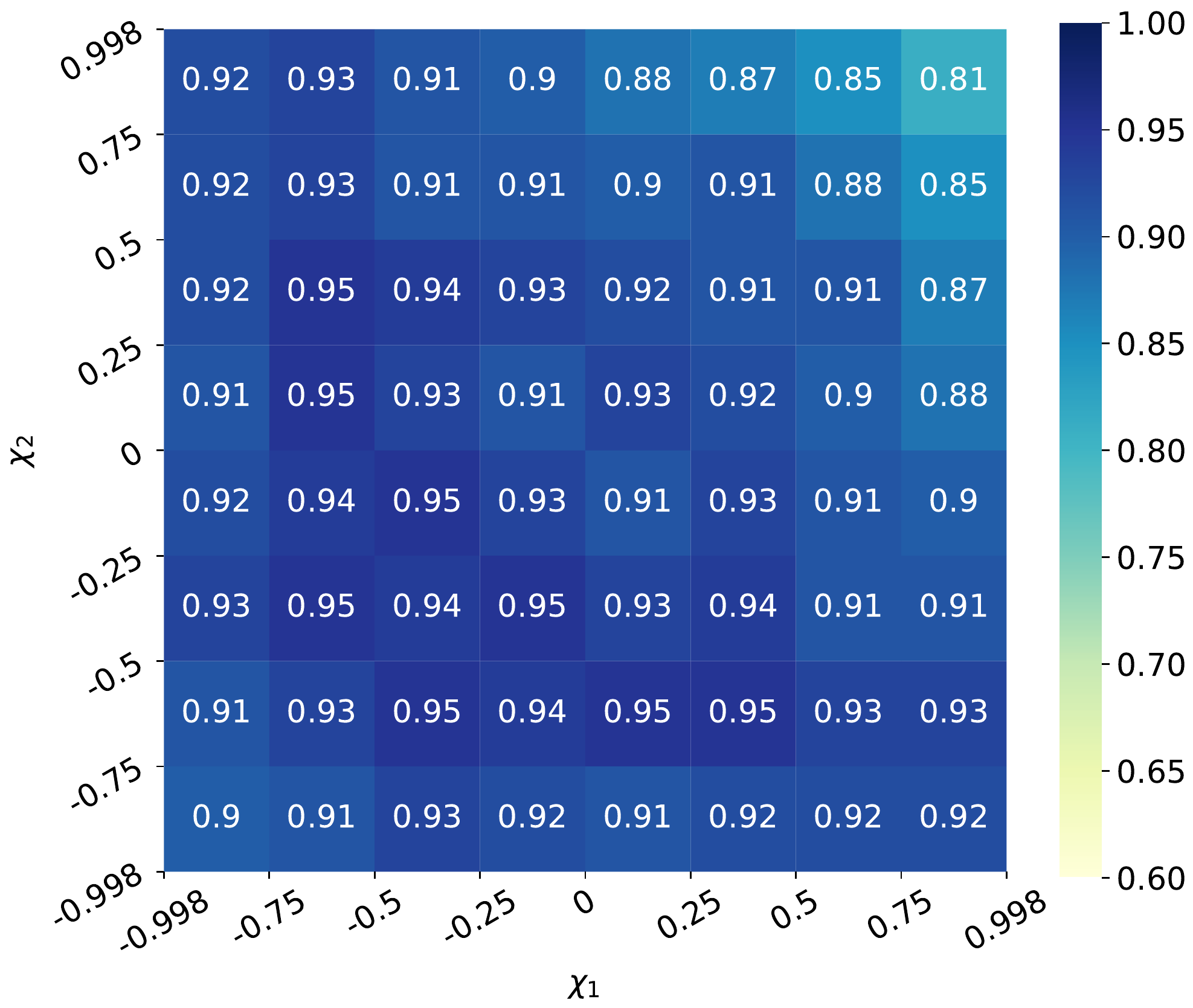}
	\caption{Model robustness on BBH masses (left) and spins (right). The colorbar and numbers are the accuracy of the model tested on different sub-datasets (in specific mass/spin ranges) of Testing Datasets.  The mirror part of $m_2 \geq m_1$ is also plotted.}
	\label{fig:model:robustness}
\end{figure*}

Based on the above study, we find that the dropout, batch normalization,
and $1\times1$ convolution techniques can improve the performance of the
basic model of CNNs~\cite{George:2016hay}. Among them, the extended model
using multiple techniques achieve better results than using a single
technique extension. In this work, ConvNet5 using dropout and batch
normalization achieves satisfactory results on both datasets. Therefore, we
will use this model in the following analysis.

%---------------------------------------------------------------------
\section{Generalization and Robustness}
\label{sec:robust}
%---------------------------------------------------------------------

In the work of \citet{George:2016hay}, who firstly introduced the CNN
structure into the field of GW data processing, {\it the generalization
ability} of deep learning algorithms in the task of GW signal detection was
discussed. Subsequent studies have pointed out that this ability is a major
advantage of deep learning algorithms over the matched
filtering~\cite{Gabbard:2017lja,Gebhard:2019ldz}. As mentioned in
Sec.~\ref{subsec:nn}, {\it generalization ability} refers to the ability of
the detection algorithm to respond to signals that are outside of the
distribution of training data~\cite{mohri:2018}. We here are the first to
study the generalization ability of the CNN model in the task of GW signal
detection, including the generalization characteristics of the CNN model in
multiple parameter ranges such as masses and aligned spins.

From the discussion in Sec.~\ref{sec:res}, we use ConvNet5 for our
following experiments, and hereinafter we refer it as ``the model''.
Considering that the discussion in Sec.~\ref{sec:res} is based only on the
performance of the model on the dataset where spins are set to zero, we
conduct a further test on the ConvNet5 model on Dataset 4.1. Dataset 4.1
introduces spin parameters on the basis of Dataset 1.2, which is used to
test the overall performance of the model in the GW detection task. After
30 epochs of training on Dataset 4.1, the model reaches an accuracy of
$93\%$, which proves that the model still has a good ability to detect GW
signals when considering the spin parameters. In the following subsections,
we will test the generalization ability of the model in masses, spins, and
the generalization ability of the nonspinning model with spinning data.

\subsection{Mass Parameters}
\label{subsec:mass}

\begin{figure*}%[htbp!]
	\centering
	\includegraphics[height = 3.4cm]{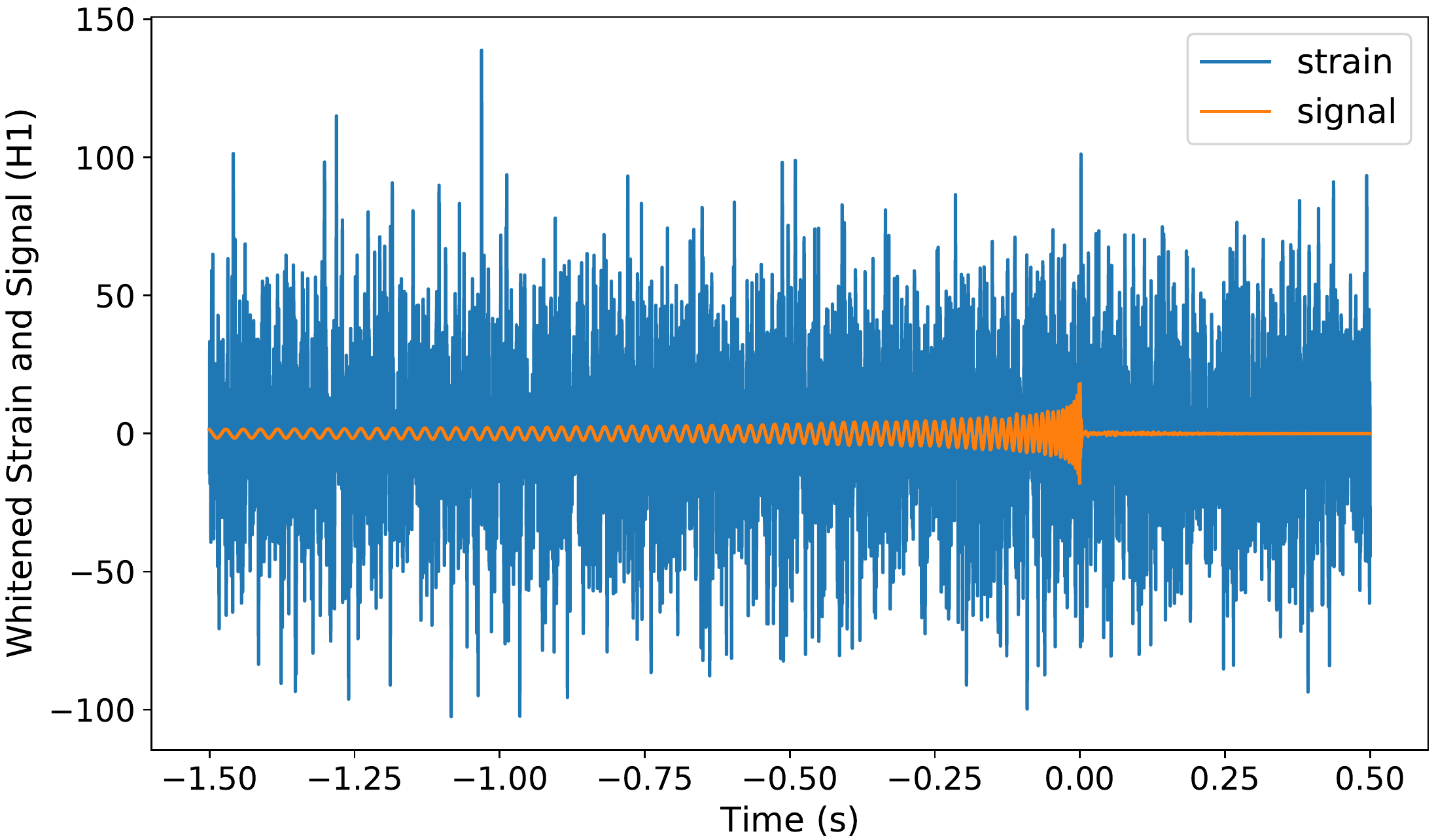}
	\includegraphics[height = 3.4cm]{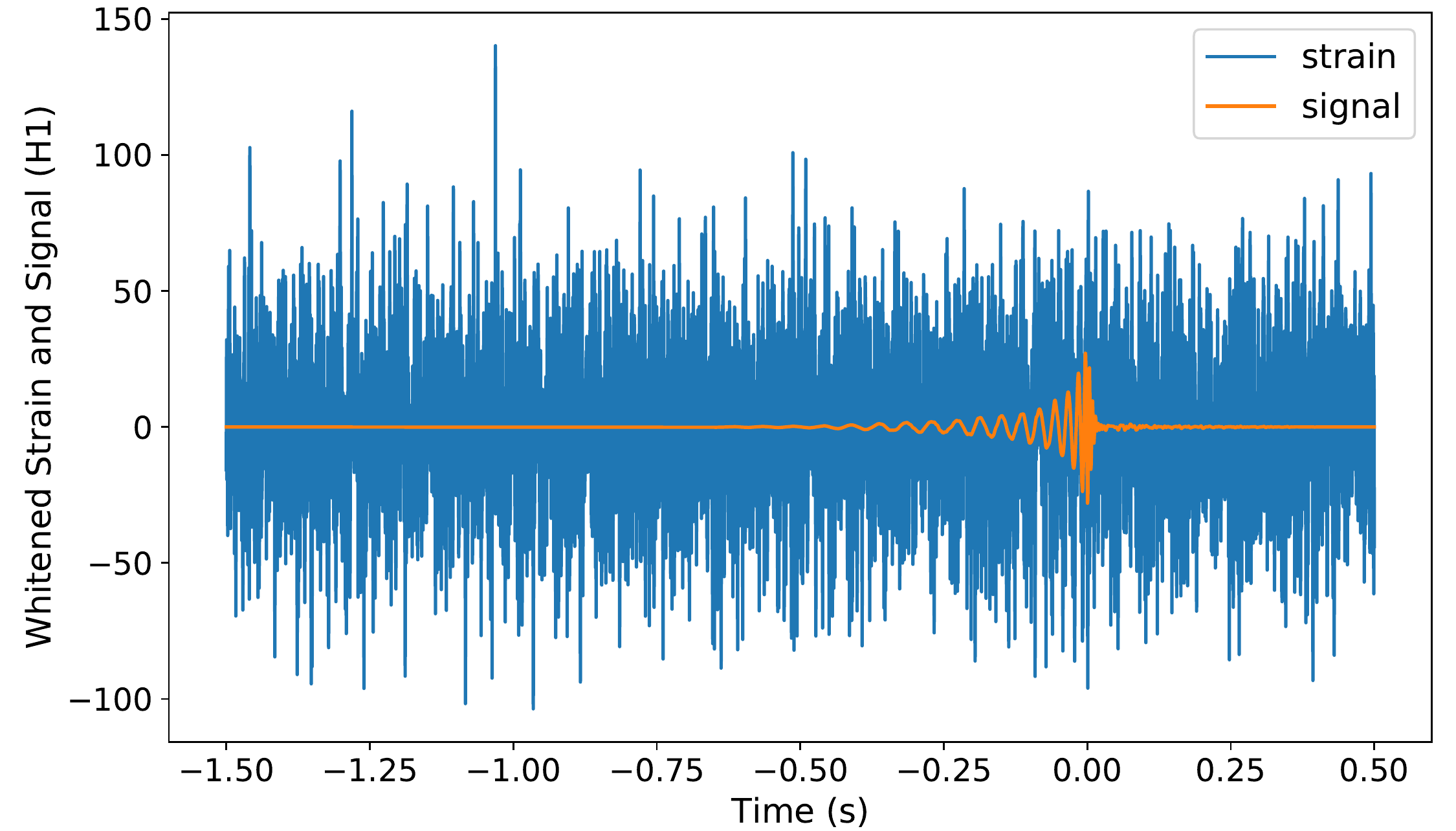}
	\includegraphics[height = 3.4cm]{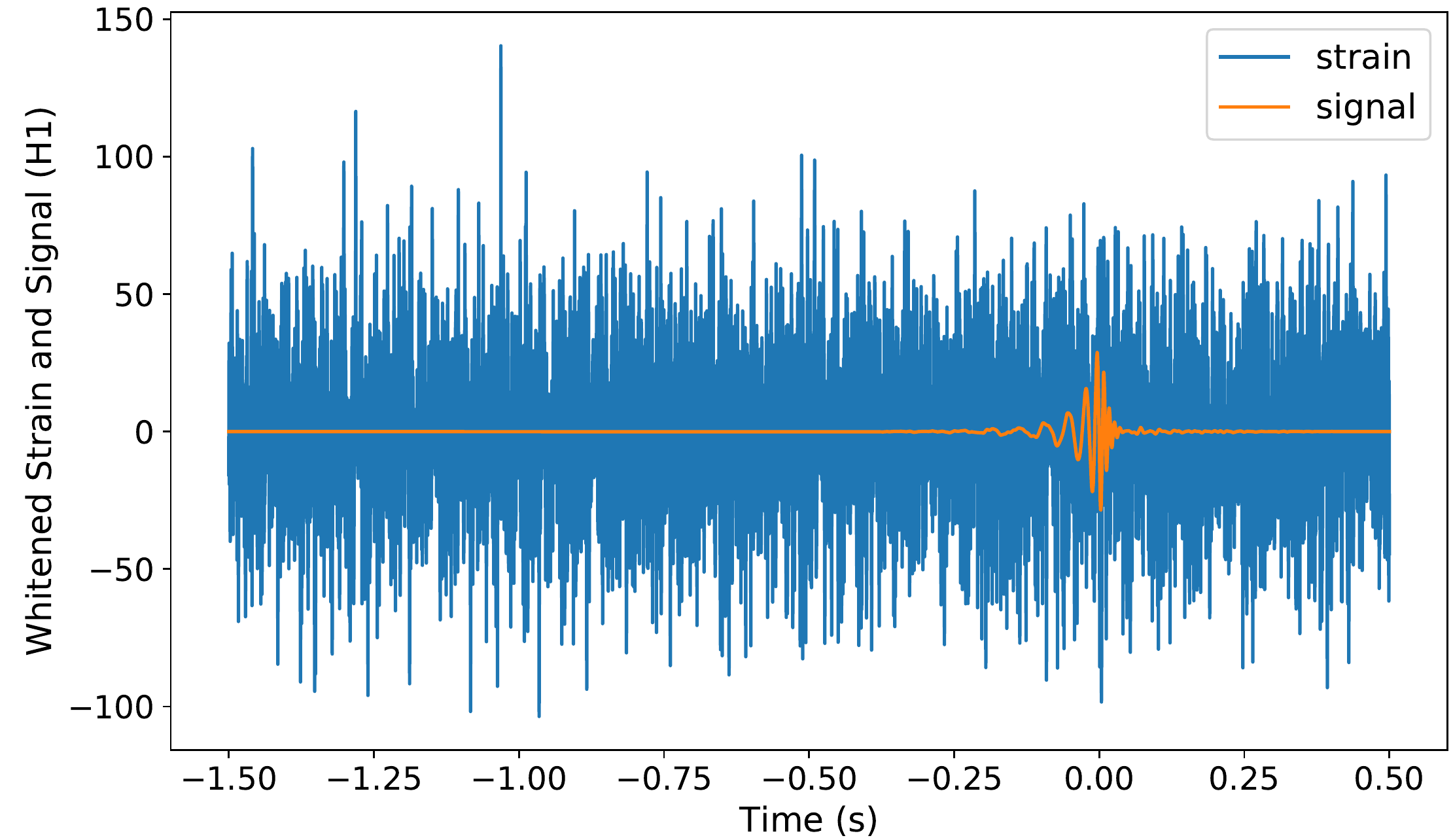}
	\includegraphics[height = 3.4cm]{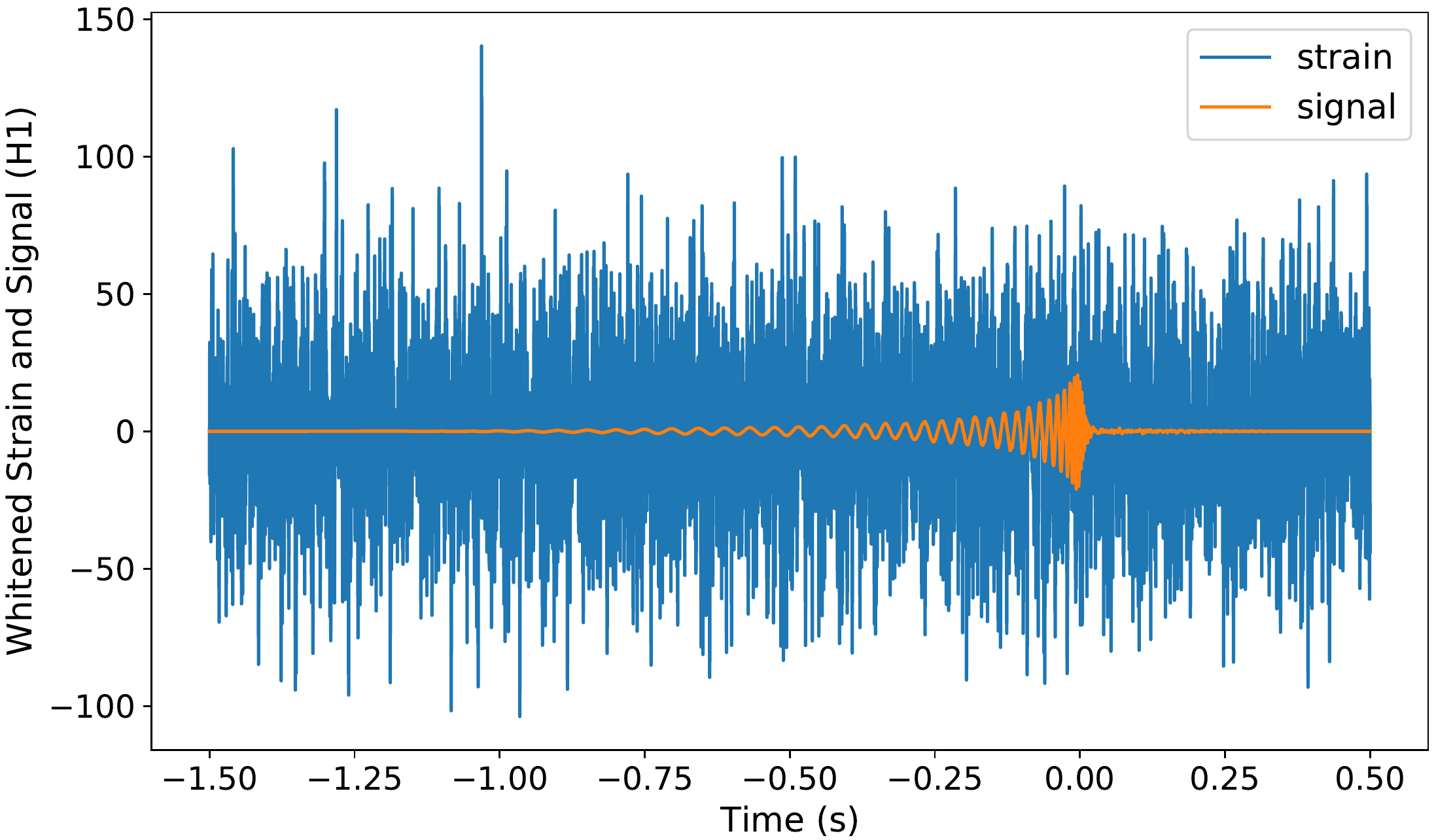}
	\includegraphics[height = 3.4cm]{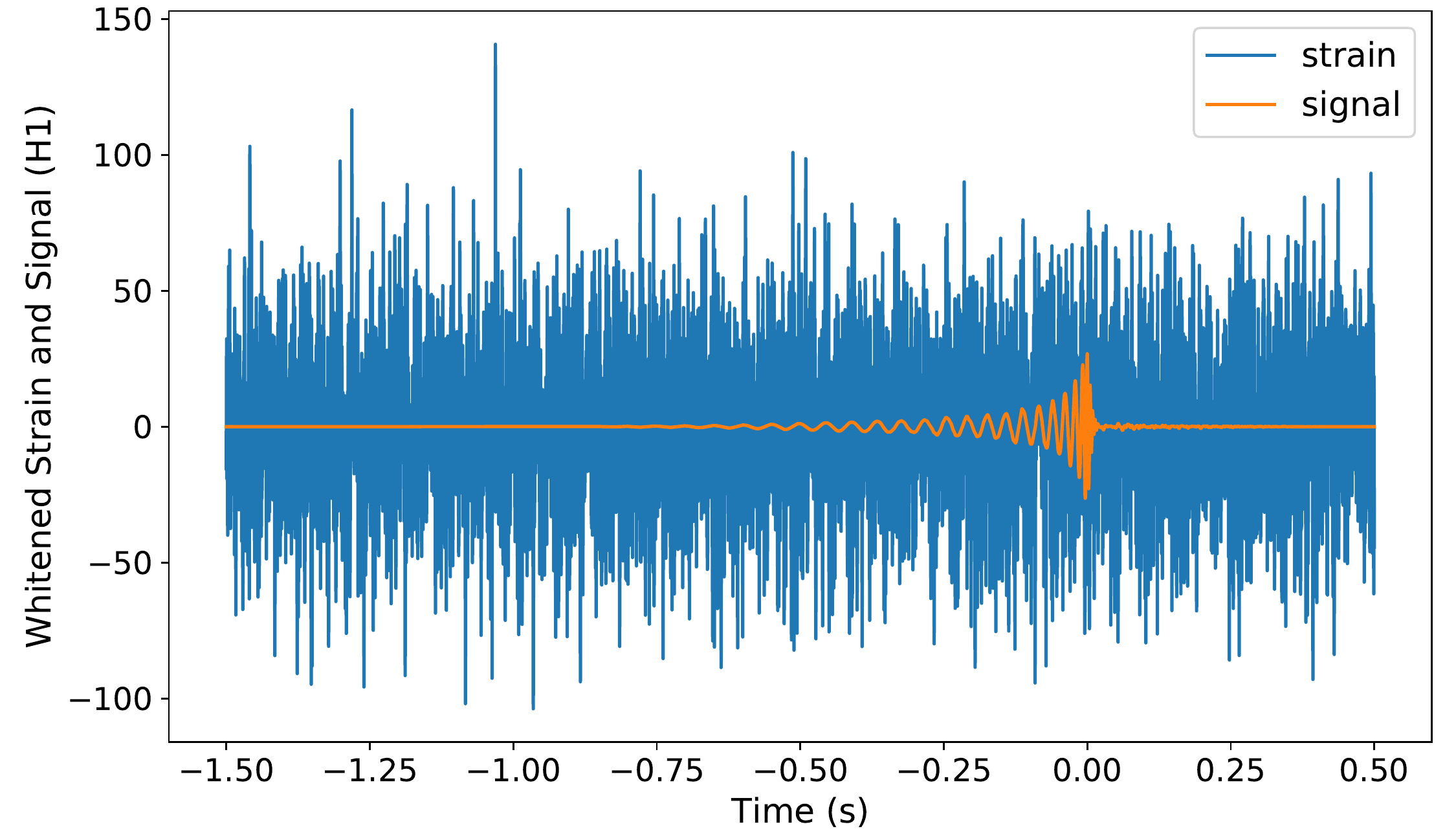}
	\includegraphics[height = 3.4cm]{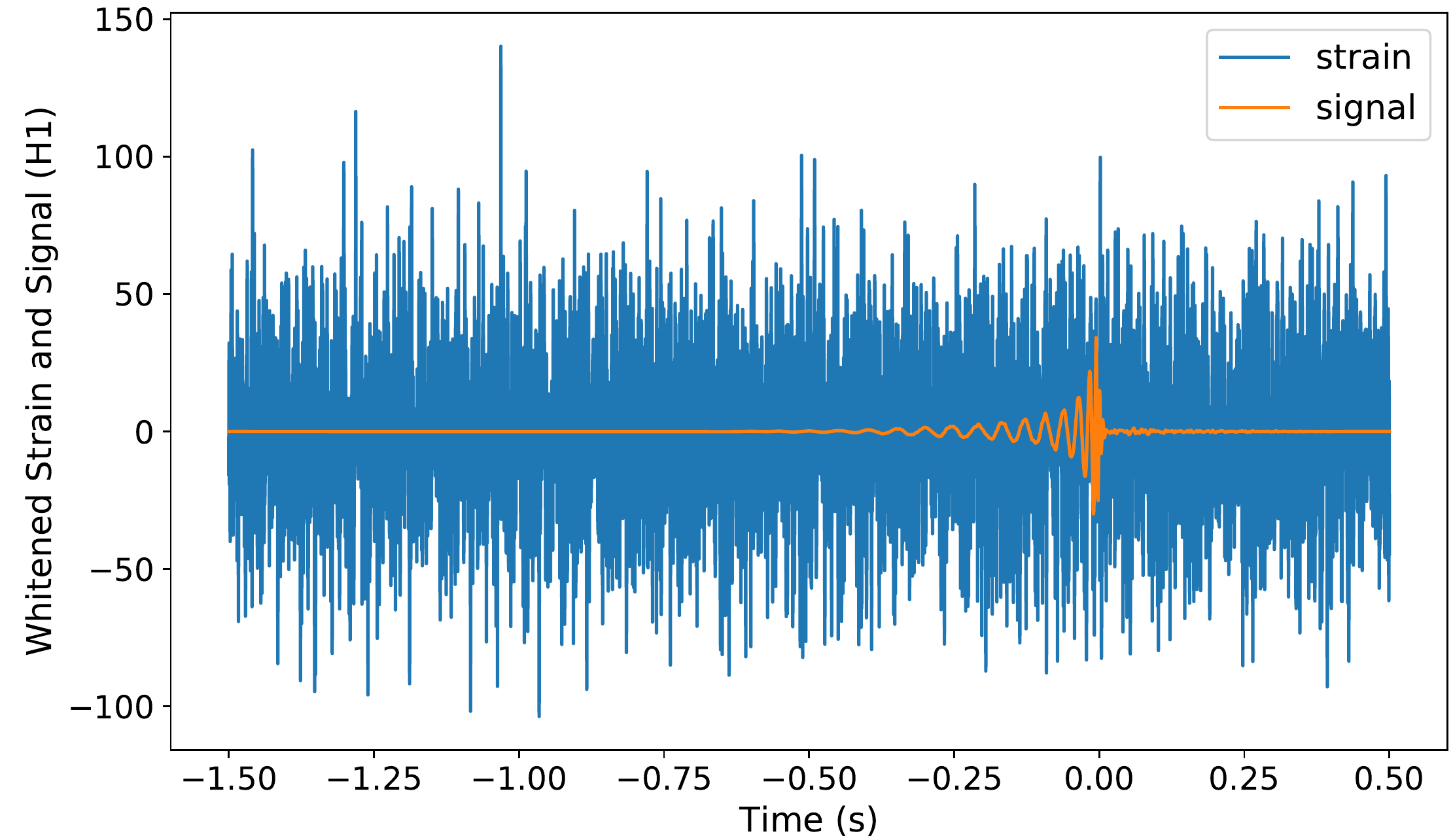}
	\caption{Simulated GW strain examples of different masses and
	spins. The upper panels are the strain examples of increasing
	masses from left to right: $(10 \, M_{\odot },10 \, M_{\odot })$,
	$(45 \, M_{\odot },45 \, M_{\odot })$, and $(80 \, M_{\odot },80 \,
	M_{\odot })$. The lower panels contain examples of different spins
	(from left to right): $(0.998, 0.998)$, $(0,0)$, and
	$(-0.998,-0.998)$.}
	\label{fig:examples:of:different:mass:spin}
\end{figure*}

To simplify the investigation, we first study the generalization ability of
the model in the GW detection task that only considers mass parameters. The
Training Dataset used in this section is Dataset 2.1 (see
Table~\ref{tab:Dataset:parameter:settings}). In this dataset, the SNR of
the GW signal is 8, the mass sampling range of the BBH is
$\left[30\,M_{\odot },60\,M_{\odot }\right]$, and the spins are set to
zero. After 30 epochs of training, the accuracy on the testing set is
$93\%$. Subsequently, we use the Testing Dataset 2 to test the model on the
sub-datasets with the mass sampling range of $\left[10\,M_{\odot
},80\,M_{\odot }\right]$. The accuracy is shown in the left panel of
Fig.~\ref{fig:model:robustness}.

We find that the model works better on sub-datasets within the original
training parameter range of $\left[30 \, M_{\odot },60 \, M_{\odot
}\right]$. For most of sub-datasets beyond the training parameter range
(i.e., outside of the range $\left[30 \, M_{\odot },60 \, M_{\odot
}\right]$), high accuracy is still achieved, indicating that the model does
have a certain generalization ability in the mass parameter. It can be seen
from the data in the figure that the model is more effective for BBHs of
high masses, and the accuracy for low-mass BBHs is lower. Studying the
signal pattern corresponding to the mass parameter, we find that this is
caused by the fact that the model is easier to identify data fluctuations
introduced by short signals. As the mass decreases, the GW signal in band
becomes longer, which is relatively closer to the characteristics of random
noise. This is harder to identify with our CNN model. We show example
waveforms of different masses and spins in
Fig.~\ref{fig:examples:of:different:mass:spin}. It is clear from the upper
panels that GW signals of smaller masses are longer, thus they are harder
to be detected in our CNN models. We leave the discussion of further
improvement of our CNN models in this parameter space to future studies.

\subsection{Spin Parameters}

Now we discuss the generalization ability of the model in spin parameters.
We use Dataset 3.1 as an illustration. The SNR of this dataset is 8, the
masses of the BBH are $30\,M_{\odot }$ and $45\,M_{\odot }$, and the range
of the spin parameter $\chi$ is $\chi_i \in [-0.5, 0.5]$ ($i=1,2$). After
30 epochs of training with the model, the accuracy on the testing set
is $92\%$. Similarly, we use Testing Dataset 3 (see
Table~\ref{tab:Testing:dataset:parameters:setting}) to test the model on
each sub-dataset with the spin parameter $\chi$ ranging in
$[-0.998,0.998]$. 

The accuracy is shown in the right panel of
Fig.~\ref{fig:model:robustness}. Similar to the generalization ability in
the mass parameter, the model performs well on the sub-datasets in the
original parameter range. It also has good accuracy on the sub-datasets
beyond the parameter range. Among them, the model performs slightly worse
on the sub-datasets with larger positive spins. This is also consistent
with the case in Sec.~\ref{subsec:mass}. As shown in the lower panels of
Fig.~\ref{fig:examples:of:different:mass:spin}, the spin parameters of the
BBHs are taken as $(0.998, 0.998)$, $(0,0)$, $(-0.998,-0.998)$,
respectively, from left to right. It is clear that with the increase of the
$\chi$ value, the GW signals become longer (the ``hang-up'' effect), thus
harder to be detected with our CNN model. Further improvements in this
aspect will be subject to future studies.

\subsection{Robustness of Nonspinning Model on Spinning Data}

What if the model is only trained with nonspinning GW signals, and it is
tested with spinning signals? We use the Dataset 1.2 (see
Table~\ref{tab:Dataset:parameter:settings}) for training that only uses the
mass parameters. After 30 epochs of training, we test the model on the
Testing Dataset 4 (see Table~\ref{tab:Testing:dataset:parameters:setting})
which contains spin parameters as well. We obtain an accuracy of
$\sim92\%$. The accuracy of the model is only slightly reduced compared to
the accuracy of $\sim94\%$ in the Dataset 1.2 where only the mass parameter
is included. It shows the generalization ability of the model that is only
trained with nonspinning GWs, but tested in data containing spin
parameters.
	
\begin{figure}%[htbp!]
		\centering
		\includegraphics[width = 8.5cm]{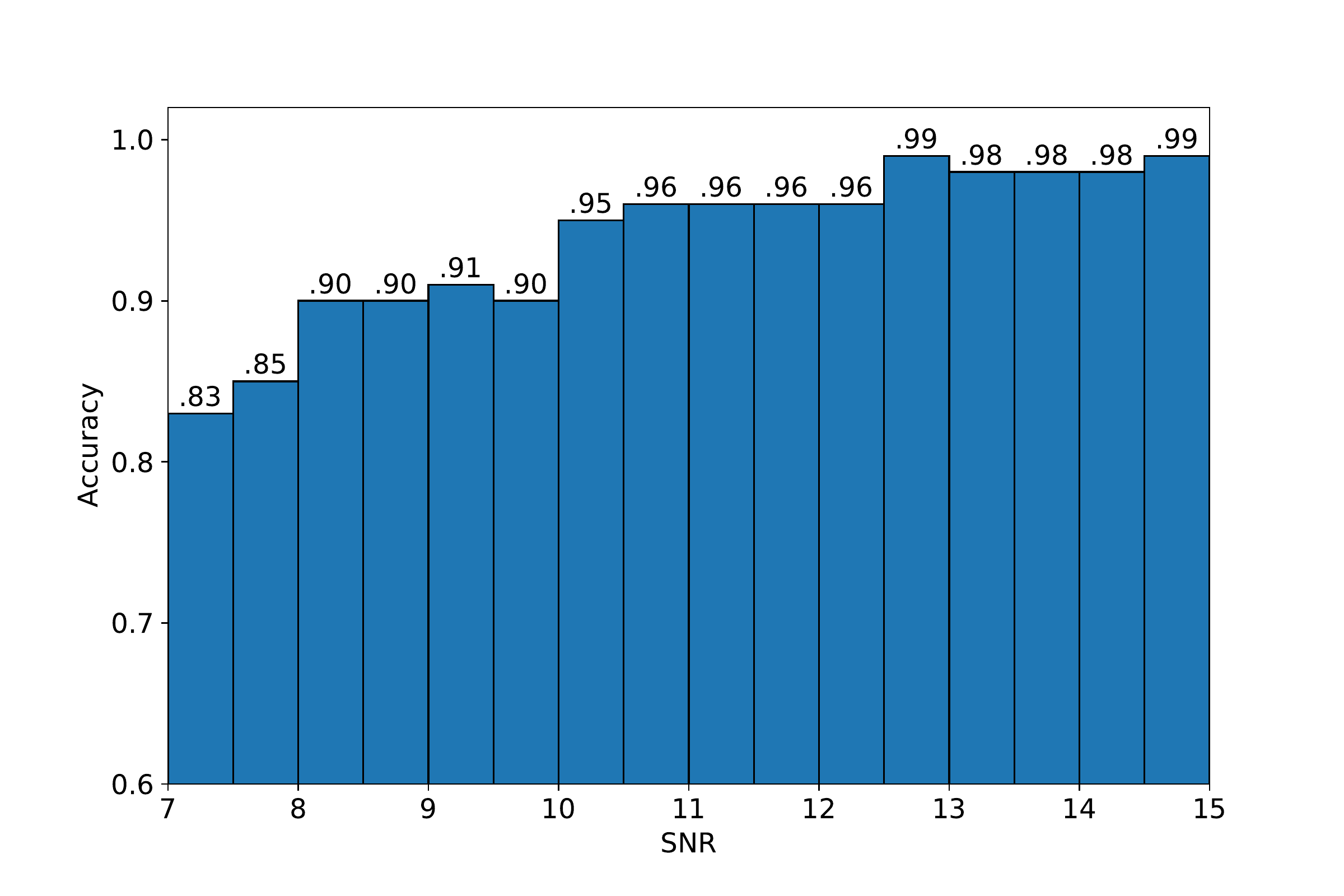}
		\caption{Robustness of CNN models that are only trained with
		nonspinning GWs on the spinning GW data for different SNRs.}
		\label{fig:mass:model:robustness:on:spin:data}
\end{figure}

We further use the Testing Dataset 5 (see
Table~\ref{tab:Testing:dataset:parameters:setting}) to test the model on
sub-datasets with different SNR distribution. The result is shown in
Fig.~\ref{fig:mass:model:robustness:on:spin:data}. As shown in the figure,
the model has the best detection efficiency on data with high SNR and it is
slightly worse on data with low SNR. However, on the sub-dataset of testing
data with SNR in the range of $[7.0, 7.5]$, the model still maintains an
accuracy of $\sim83\%$. This shows that the CNN model which was trained
only with nonspinning data is robust to data containing nonzero spin
parameters.

\subsection{Conclusion}

From the above study, it can be seen that the CNN model shows good
generalization ability for data of different parameter ranges, which is a
major advantage over the matched filtering method in using the CNN model.
The matched filtering method is based on the existing template bank. Its
search and response to GW signals are limited to the existing waveforms.
The GW signals beyond the existing waveforms cannot be easily detected. The
generalization ability of the CNN model in the task of GW signal detection
will help us to discover signals beyond the existing templates. It may also
be able to detect signals that imprint orbital
eccentricity~\cite{Liu:2019jpg}, orbital precession~\cite{Babak:2016tgq},
and deviations from general relativity~\cite{Shao:2017gwu}. Such a
generalization ability will undoubtedly play an essential role in searches
of GW signals beyond what we have in the template bank. More studies are
needed along this direction.

%---------------------------------------------------------------------
\section{Discussion}
\label{sec:disc}
%---------------------------------------------------------------------

For the first time, this work specifically studied the effects on CNN
models brought by the improving techniques---such as dropout, batch
normalization, and $1\times1$ convolution---in the task of GW signal
detection. Comparisons of models with the combination of different
improving techniques were made. Our simplified experiments show that
dropout, batch normalization, and the $1\times1$ convolution can
significantly improve the accuracy of the basic
model~\cite{George:2016hay}. In addition, we conducted specific experiments
and discussed the generalization ability of CNNs in GW signal detection
tasks. Experiments show that, the CNN model has good generalization ability
over multiple parameter ranges, including masses and aligned spins. It is a
major advantage of the deep learning models compared with the
matched-filtering method.

In recent years, with the rapid development in the field of deep learning,
deep learning models and algorithms have emerged one after another. CNN is
only a basic model in the market of deep learning though, it has shown
excellent performance in many fields such as image recognition and signal
detection~\cite{Goodfellow:2016}. As the difficulty of the task continues
to increase, the requirements for feature extraction and processing are
getting higher and higher. The network depth of the CNN is limited by the
gradient propagation problem, resulting in a limitation on its
performance~\cite{Goodfellow:2016}. The proposal of the residual network
structure greatly alleviates the problem of gradient propagation in CNNs
and makes the depth limit of CNN models rapidly expand~\cite{He:2015wrn}.
In the field of time series processing, the ``Long Short Term Memory''
model is widely used in multiple tasks such as speech processing and stock
price prediction, and has achieved excellent results~\cite{Jurgen:1997}. At
present, some studies have introduced the above structure into the process
of GW data processing~\cite{Dreissigacker:2019edy, Shen:2017jkj}. However,
in the task of GW signal detection, there is still a lack of unified
investigation on these models. Comparing these models (including CNN
models) on GW signal detection tasks, such as the recognition rate, model
size, running time and other indicators, is crucial to find an optimal
model structure suitable for real-world task. It is a worthwhile direction
for future studies.

For the sake of simplifying the investigation, this work only uses
simulated white noise to construct GW sample data. In order to make the
model more in line with the characteristics of reality detection, real
noise construction from aLIGO/Virgo data can be conducted. At the same
time, the data used in this article are only for experimental purposes, so
the amount of data is limited. By increasing the number of samples in the
dataset, the performance of the model can be further improved. Ensemble
learning is also an optimization technique that is worth considering, where
the basic idea is to train multiple neural network models, and to use
voting and other common decision-making strategies in order to make up for
the decision-making defects of a single model and to improve the
decision-making ability of the overall model~\cite{Opitz:1999}. These
aspects deserve detailed studies on their own.

%---------------------------------------------------------------------

%---------------------------------------------------------------------
\acknowledgments

This work was supported by the National Natural Science Foundation of China
(11975027, 11991053, 11690023, 11721303), the Young Elite Scientists Sponsorship
Program by the China Association for Science and Technology (2018QNRC001),
the Max Planck Partner Group Program funded by the Max Planck Society, and
the High-performance Computing Platform of Peking University. It was
partially supported by the Strategic Priority Research Program of the
Chinese Academy of Sciences through the Grant No. XDB23010200.

%---------------------------------------------------------------------
\bibliography{refs}
%---------------------------------------------------------------------

\end{document}